\documentclass[journal,draftclsnofoot,onecolumn]{IEEEtran}
\usepackage{cite}
\usepackage[pdftex]{graphicx}
	\ifCLASSINFOpdf
	 	\usepackage[pdftex]{graphicx}
  	\else
	\fi
\usepackage[cmex10]{amsmath}
\usepackage{amsthm}
\usepackage{array}
\usepackage{color}

\usepackage{amssymb}

\begin{document}

\title{A New and More General Capacity Theorem for the Gaussian Channel with Two-sided Input-Noise Dependent State Information}


\author{Nima~S.~Anzabi-Nezhad,
        Ghosheh~Abed~Hodtani,~and~Mohammad~Molavi~Kakhki
\thanks{N. S. Anzabi-Nezhad is with the Department of Electrical Engineering, Ferdowsi University of Mashhad, Iran, email: nima.anzabi@gmail.com}
\thanks{G. A. Hodtani is with the Department of Electrical Engineering, Ferdowsi University of Mashhad, Iran, email: ghodtani@gmail.com}
\thanks{M. Molavi Kakhki is with the Department of Electrical Engineering, Ferdowsi University of Mashhad, Iran, email: molavi@um.ac.ir}}

\maketitle

\begin{abstract}
In this paper, a new and general version of Gaussian channel in presence of two-sided state information correlated to the channel input and noise is considered. Determining a general achievable rate for the channel and obtaining the capacity in a non-limiting case, we try to analyze and solve the Gaussian version of the Cover-Chiang theorem -as an open problem- mathematically and information-theoretically.
Our capacity theorem, while including all previous theorems as its special cases, explains situations that can not be analyzed by them; for example, the effect of the correlation between the side information and the channel input on the capacity of the channel that can not be analyzed with Costa's ``writing on dirty paper" theorem. Meanwhile, we try to introduce our new idea, i.e., describing the concept of ``cognition" of a communicating object (transmitter, receiver, relay and so on) on some variable (channel noise, interference and so on) with the information-theoretic concept of ``side information" correlated to that variable and known by the object. According to our theorem, the channel capacity is an increasing function of the mutual information of the side information and the channel noise. Therefore our channel and its capacity theorem exemplify the ``cognition" of the transmitter and receiver on the channel noise based on the new description. Our capacity theorem has interesting interpretations originated from this new idea. \end{abstract}

\IEEEpeerreviewmaketitle


\section{Introduction} \label{sec.introduction} 
Side information channel has been actively studied since its initiation by Shannon \cite{shannon}. Coding for computer memories with defective cells was studied by Kusnetsov-Tsybakov \cite{kusnetsov}. Gel'fand-Pinsker (GP) \cite{GP} determined the capacity of channels with channel side information (CSI) known non-causally at the transmitter. Heegard-El Gamal \cite{HG} obtained the capacity when the CSI is known only at the receiver. Cover-Chiang \cite{coverchiang} extended these results to a general case where correlated two-sided state information are available at the transmitter and at the receiver. Costa \cite{costa} obtained an interesting result by carefully investigating the GP theorem for the Gaussian channel, i.e., he proved that the capacity of the Gaussian channel with an interference known at the transmitter is the same as the capacity of interference free channels.
There are many other important researches in the literature, e.g.\cite{sajafar,keshet2008,merhav2007}. The results for the single user channel have been generalized possibly to multi user channels, at least in special cases \cite{steinberg2005,sigurjonsson2005,kim2004,khosravi2010,philosof2009,steinberg20052}.\\

\subsection*{Our Motivations}
In this paper, we focus on the Gaussian channel in presence of side information for two major aims: First, analyzing the problem of capacity of the Gaussian channel in presence of two sided state information -the Gaussian version of Cover-Chiang theorem \cite{coverchiang}, mathematically and information-theoretically.  Second we try to present an information-theoretical description of the concept of ``cognition" of the transmitter and or receiver in an improved manner.\\

\subsubsection*{First motivation} 
\begin{figure}[!t]
\centering
\includegraphics[width=3.5in]{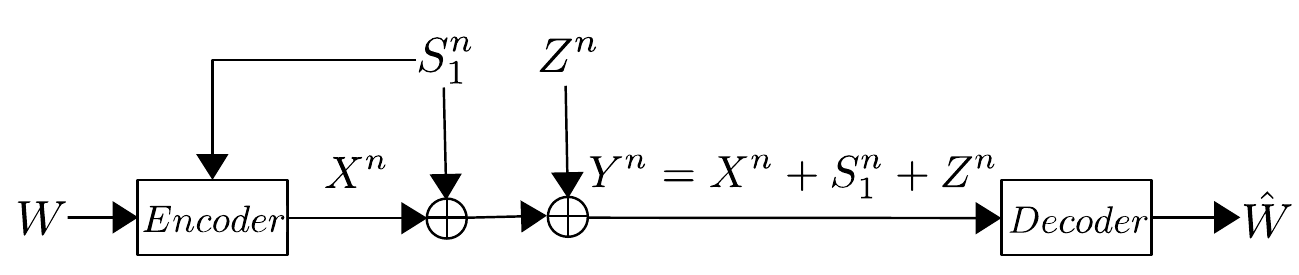}
\caption{Gaussian channel with additive interference known non-causally at the transmitter.} 
\label{figure1}
\end{figure}

In this paper, we try to analyze the Gaussian version of the Cover-Chaing unifying theorem \cite{coverchiang}. The problem of the effect of side information at the transmitter in a Gaussian channel, in a special case, first, has been studied in Costa's "writing on dirty paper" \cite{costa}. Let us consider a Gaussian channel with side information known non-causally at the transmitter as depicted in Fig. \ref{figure1}. We denote the side information at the transmitter, the channel input, the channel output, the channel noise and the auxiliary random variable at the transmitter by $S_{1}$, $ X $, $ Y $, $Z$ and $ U $, respectively. Moreover, it is assumed that $S_{1}$ and $Z$ are Gaussian random variables with powers $Q_{1}$ and $N$ respectively and $X$ has the power constraint $ E\left\lbrace X^{2} \right\rbrace \leq P $.

Costa \cite{costa} shows that the capacity of this channel is surprisingly the same as the capacity of the channel without side information. An important assumption in Costa theorem is that in the definition of the channel, there is no restriction for the correlation  between $X$ and $S_{1}$. However, Costa shows that the maximum rate is obtained when $X$ and $S_{1}$ are independent and $U$ is a linear function of $X$ and $S_{1}$. Hence, his theorem is only applicable to cases where $X$ and $S_{1}$ have the chance to be uncorrelated. Therefore a theorem which can handle the capacity of Gaussian channels when there exists a specific correlation between $X$ and $S_{1}$ is theoretically and practically important. One example for correlated input and side information is cognitive interference channels in which the transmitted sequence of one transmitter is a known interference for the other transmitter and these two sequences may be dependent to each other. Another example is a measurement system where the measuring signal may affect the system under measurement. This is equivalent to an interfering signal which is dependent on the original measuring signal.

\begin{figure}[!t]
\centering
\includegraphics[width=3.5in]{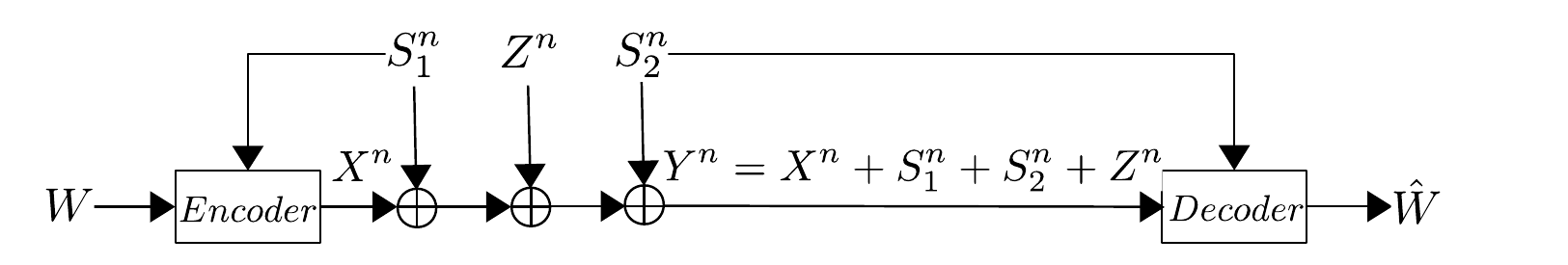}
\caption{Gaussian channel with correlated side information known non-causally at the transmitter and at the receiver.} 
\label{figure2}
\end{figure}

Another related question is about the side information $S_{2}$ known non-causally at the receiver (if exists as in Fig. \ref{figure2}). The question now arises is that: How does the receiver knowledge $S_{2}$, correlated to $\left (X,S_{1}\right )$ affect the channel capacity? And how much does the receiver information about $X$ and $S_{1}$, available through $S_{2}$, change the channel capacity?

Some communication scenarios in which the channel input and the side information may be correlated and the related investigations can be found in \cite{merhav2007} and \cite{Yu-Chih2012}. In \cite{merhav2007} the problem of optimum transmission rate under the requirement of minimum mutual information $I\left (S_{1}^{n};Y^{n}\right )$ is investigated. Moreover both \cite{merhav2007} and \cite{Yu-Chih2012} study Costa's ``writing on dirty paper" problem where the side information is correlated to the input of the channel (our motivation), when only side information known at the transmitter exists. We, in another work, have considered and solved the problem of the capacity of Gaussian channel with two-sided state information in a limited case \cite{AnzabiHodtani2013CommLett}. 

Moreover, examining the Gaussian channel with two-sided state information with dependency on the channel noise and channel input, we try to solve the Gaussian version of Cover-Chiang theorem \cite{coverchiang} as an open problem.   \\

\subsubsection*{Second motivation} One of the most known and important applications of the channels with side information is information theoretically describing the concept of ``cognition" of the transmitter in communication scenarios. Side information in this description, for example,  may be the interference which transmitter exactly knows all about it. Two questions arise about this description:

1) It is usually expected the knowledge about or cognition on something to be ``quantitative". For example the cognition that the transmitter can acquire about the interference may be incomplete or partial. So one question is: How can we describe the ``quantity" or ``amount" of the transmitter cognition? The investigations of the channels with partial CSI try to answer this question, for example \cite{rosenzweig2005,chen2001,zaidi2006,Gueguen2009,bahmani2012}.

2) It is possible in a communication scenario that the transmitter has knowledge about more than one variable in the channel. For example in a cognitive interference channel the transmitter may have knowledge about the interference originated by the other transmitter and at the same time about the channel noise. Hence, the other question is: How can we describe the ``cognition  that the transmitter  has on some variables".

In this paper, we propose  describing the concept of the ``transmitter and or receiver cognition on some variables" by side information available at the transmitter and or receiver \textit{probabilistically dependent} on those variables. Hence, the side information known at the transmitter correlated to the variable $A$, describes the transmitter cognition on $A$ and the amount of this cognition increases as the correlation between the side information and the variable $A$ increases.
Distinguishing between this meaning of ``cognition" from the usual meaning widely used in the literature, it may be proper to use the word ``\textbf{re-cognition} (of the transmitter or receiver on something)" for it.

Hence in a Gaussian channel in presence of two-sided state information depicted in Fig. \ref{figure2},  $S_{1}$ which is the side information known at the transmitter can be interpreted as the transmitter re-cognition on the channel noise, if $S_{1}$ is correlated with $Z$. 
It is seen that our first motivation, not only can be seen as an effort to solve an important open problem, but also ,if solved, it can exemplify this new description.\\

\subsection*{Our Work} To provide the above motivations, we define a Gaussian channel in presence of two-sided state information where the channel input $X$, side information $(S_{1},S_{2})$ and the channel noise $Z$ are arbitrarily correlated. Using the extended version of Cover-Chiang unifying theorem \cite{coverchiang} to continuous alphabets, we prove a general achievable rate for the channel (lemma 1).
Then, we obtain a general upper bound for the channel in the case that the channel input $X$, the side information $\left (S_{1},S_{2} \right )$ and the channel noise $Z$, form the Markov chain $X \rightarrow \left (S_{1},S_{2}\right )\rightarrow Z$ (lemma 2) and we show the coincidence of the lower and upper bounds under this circumstance and therefore establish our capacity theorem for the channel. Using our probabilistic description of ``re-cognition" of the transmitter, this circumstance can be explained as follows: if the whole ``re-cognition" that  the transmitter has got on the channel noise, is gained from the side information $\left ( S_{1},S_{2}\right )$ -that is a meaningful and practically acceptable circumstance in our communication scenario- then the Markov chain $X \rightarrow \left (S_{1},S_{2}\right )\rightarrow Z$ must be satisfied. The obtained channel capacity can be expressed as an increasing function of the mutual information between the side information $\left (S_{1},S_{2}\right )$ and the channel noise $Z$ (i.e. $I\left (S_{1}S_{2};Z\right )$ ) and this shows that our new description of ``re-cognition" of the transmitter and the receiver can be exemplified by our channel and its capacity.
\subsection*{Paper Organization} This paper is organized as follows: in section II, we briefly review the Cover-Chiang and the Gel'fand-Pinsker theorems and then introduce a scrutiny of the Costa theorem. In section III, we define our Gaussian channel thoroughly and prove a general lower bound for the defined channel and then obtain a general upper bound for the channel in mentioned case, which coincides with the lower bound and hence is the capacity of the channel. In Section IV, we examine the proved capacity in special cases and interpret them. Specifically, we explain that how this capacity theorem can exemplify the new description of the ``re-cognition" of transmitter and or receiver on something. Section VI contains the conclusion. The proofs of lower and upper bounds of the capacity of channel and two lemmas used in our proofs are given in the Appendix.

\section{A Review of Previous Related Works}\label{sec.review} 
To clarify our approach in subsequent sections, in this section we first briefly review the Cover-Chiang capacity theorem for channels with side information available at the transmitter and at the receiver. We then review the Gel'fand-Pinsker (GP) theorem which is a special case of Cover-Chiang theorem  when side information is known only at the transmitter. Finally Costa theorem (‍‍‍‍‍``writing on dirty paper" theorem), which is the Gaussian version of the GP theorem, is deeply investigated. 
\subsection{Cover-Chiang Theorem}
\label{sec.II.Cover-Chiang Theorem}
\begin{figure}[!t]
\centering
\includegraphics[width=3.5in]{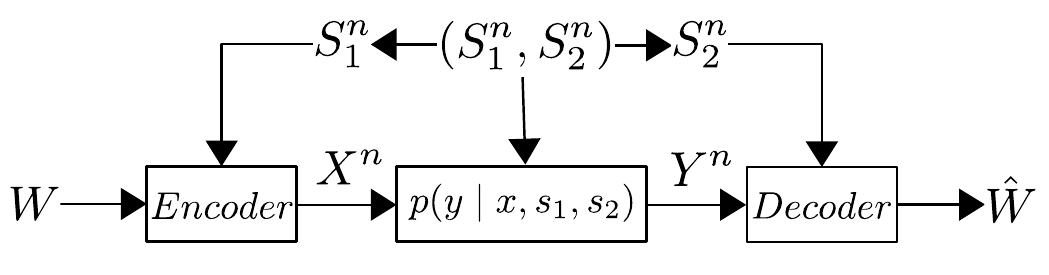}
\caption{Channel with side information available non-causally at the transmitter and at the receiver.}
\label{figure3}
\end{figure}

 Fig. \ref{figure3} shows a channel with side information known at the transmitter and at the receiver where $ X^{n} $ and $ Y^{n} $ are the transmitted and the received sequences respectively. The sequences $ S_{1}^{n} $ and $ S_{2}^{n} $ are the side information known non-causally at the transmitter and at the receiver respectively. The transition probability of the channel, $ p\left( y\mid x,s_{1},s_{2}\right)  $, depends on the input $ X $, the side information $ S_{1} $ and $ S_{2} $. It can be shown that if the channel is memoryless and the sequences $ \left(S_{1}^{n},S_{2}^{n}\right)  $ is  independent and identically distributed (i.i.d.) random variables under $ p\left(s_{1},s_{2} \right)  $, then the capacity of the channel is \cite{coverchiang}:
\begin{equation}
C=\max_{p\left(u,x\mid s_{1} \right) }\left[I\left(U;S_{2},Y \right)-I\left( U;S_{1}\right)   \right] \label{eq.1} 
\end{equation}
where the maximum is over all distributions:
\begin{equation}
p\left(y,x,u,s_{1},s_{2}\right) =p\left(y\mid x,s_{1},s_{2} \right)p\left(u,x\mid s_{1} \right)p\left(s_{1},s_{2} \right) \label{eq.2}  
\end{equation}
and $ U $ is an auxiliary random variable.

It is important to note that the Markov chains:
\begin{equation}
S_{2}\longrightarrow S_{1}\longrightarrow UX \label{eq.3} 
\end{equation}
\begin{equation}
U\rightarrow XS_{1}S_{2}\rightarrow Y \label{eq.3+1} 
\end{equation}
are satisfied for all distributions in (\ref{eq.2}).

\subsection{Gel'fand-Pinsker (GP) Theorem}
\label{sec.II.Gel'fand-Pinsker Theorem}
\begin{figure}[!t]
\centering
\includegraphics[width=3.5in]{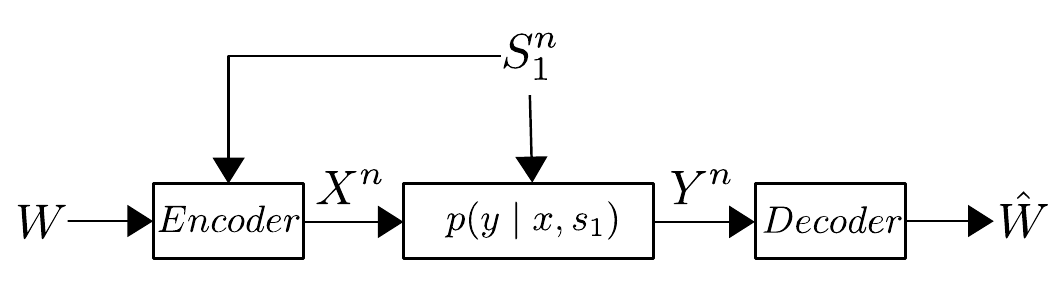}
\caption{Channel with side information known at the transmitter.}
\label{figure4}
\end{figure}

This theorem is special case of Cover-Chiang theorem when $ S_{2}=\phi $. According to GP theorem \cite{GP}:

A memoryless channel with transition probability $ p\left(y \mid x,s_{1} \right)  $ and side information sequence $ S_{1}^{n} $ i.i.d. with $ p\left(s_{1} \right)  $ known non-causally at the transmitter depicted in Fig. \ref{figure4} has the capacity
\begin{equation}
C=\max_{p\left(u,x \mid s_{1} \right) } \left[I\left(U;Y \right) -I\left(U;S_{1} \right)  \right] \label{eq.4} 
\end{equation}
for all distributions:
\begin{equation}
p\left(y,x,u,s_{1}\right) =p\left(y\mid x,s_{1} \right)p\left(u,x\mid s_{1} \right)p\left(s_{1}\right) \label{eq.5}  
\end{equation}
where $ U $ is an auxiliary random variable.

\subsection{Costa's ``Writing on Dirty Paper"}
\label{sec.II.Costa's Writing on Dirty Paper}
Costa \cite{costa} examined the Gaussian version of the channel with side information known at the transmitter (Fig. \ref{figure1}).
As can be seen, the side information is considered as an additive interference at the receiver. Costa showed that the channel, surprisingly, has the capacity $ \frac{1}{2}\log\left(1+\frac{P}{N}\right)  $, which is the the same for channels with no interference $ S_{1} $. Costa derived this capacity by using the results of Gel’fand-Pinsker theorem extended to random variables with continuous alphabets. In this subsection, we first introduce the Costa assumptions and then present a proof for this theorem in such a way that it enables us to introduce our channel and develop our theorem in subsequent sections.\\

The channel is specified  with properties C.1-C.3 below:

\paragraph*{\textbf{C.1}}
\quad $ S_{1}^{n} $ is a sequence of Gaussian i.i.d. random variables with distribution $ S_{1}\sim \mathcal{N}\left(0,Q_{1} \right)  $.

\paragraph*{\textbf{C.2}}
\quad The transmitted sequence $ X^{n} $ is assumed to have the power constraint $ E\left\lbrace X^{2} \right\rbrace \leq P $.

\paragraph*{\textbf{C.3}}
\quad The output is given by $ Y^{n}=X^{n}+S_{1}^{n}+Z^{n} $, where $Z^{n} $ is the sequence of white Gaussian noise with zero mean and power $ N $ i.e. $ Z\sim \mathcal{N}\left(0,N \right)  $  and independent of $ (X,S_{1}) $. The sequence $S_{1}^{n}$ is non-causally known at the transmitter.

It is readily seen that the distributions $ p\left(y,x,u,s_{1}\right) $ having the above three properties are in the form of (\ref{eq.5}). We denote the set of all these $ p\left(y,x,u,s_{1}\right) $'s with $\mathcal{P}_{C}$.
Although for the Costa channel described above, no restriction has been imposed on the correlation between $X$ and $S_{1}$, in Costa theorem, the maximum rate corresponds to independent $X$ and $S_{1}$, and $U$ in form of linear combination of $X$ and $S_{1}$. We define $\mathcal{P}'_{C}$ as a subset of $\mathcal{P}_{C}$ with elements $p'\left (y,x,u,s_{1} \right )$ having the following properties as well as properties C.1-C.3 mentioned before: 

\paragraph*{C.4}
\quad $X$ is a zero mean Gaussian random variable with the maximum average power $P$ and \textit{independent} of  $ S_{1}$.

\paragraph*{C.5}
\quad The auxiliary random variable $U$ takes the linear form $ U=\alpha\ S_{1}+X $.\\

It is clear that the set $\mathcal{P}'_{C}$ (described in C.1-C.5) and their marginal and conditional distributions are subsets of corresponding $\mathcal{P}_{C}$'s (described in C.1-C.3).\\ 
  
\textit{Achievable rate for Costa channel}: From (\ref{eq.4}), when extended to memoryless channels with discrete time and continuous alphabets, we can obtain an achievable rate for the channel.

The capacity of Costa channel can be written as:
\begin{equation}
C_{Costa}=\max_{p\left( u,x \mid s_{1}\right) }\left[I\left(U;Y \right) -I\left(U;S_{1} \right)  \right]\label{eq.6}
\end{equation}
where the maximum is over all $p\left (y,x,u,s_{1} \right )$'s in $\mathcal{P}_{C}$. Since $\mathcal{P}'_{C}\subseteq \mathcal{P}_{C}$ we have:
\setlength{\arraycolsep}{0.0em}
\begin{eqnarray}
C_{Costa}&{}\geq{}& \max_{p'\left( u,x \mid s_{1}\right) }\left[I\left(U;Y \right) -I\left(U;S_{1} \right)  \right]\label{eq.7}\\
&{}={}&\max_{p'\left( u \mid x,s_{1}\right) p'\left( x \mid s_{1}\right)  }\left[I\left(U;Y \right) -I\left(U;S_{1} \right)  \right]\label{eq.8}\\
&{}={}&\max_\alpha\left[I\left(U;Y \right) -I\left(U;S_{1} \right)  \right]\label{eq.9}
\end{eqnarray}
\setlength{\arraycolsep}{5pt}
The expression in the last bracket is calculated for distributions $ p'\left(y,x,u,s_{1}\right) $ in $\mathcal{P}'_{C}$ described in C.1-C.5. Thus, defining $ R\left ( \alpha \right ) = I\left(U;Y \right) -I\left(U;S_{1} \right) $, $ \max_{\alpha} R\left(\alpha \right) $ is an achievable rate for the channel. $R\left(\alpha \right)$ and $\max_{\alpha}R\left(\alpha \right)$ is calculated as:
\begin{equation}
R \left(\alpha \right)= \dfrac {1}{2}\log\left( \dfrac {P\left(P+Q_{1}+N \right) }{PQ_{1}\left(1-\alpha \right)^{2}+N\left(P+\alpha^{2}Q_{1} \right)  } \right), \label{eq.10}
\end{equation}
and
\begin{equation}
\max_{\alpha}R\left(\alpha \right)=R\left(\alpha^{\ast} \right)=\frac{1}{2}\log\left(1+\frac{P}{N} \right) \label{eq.11} 
\end{equation}
where
\begin{equation}
\alpha^{\ast}=\frac{P}{P+N}.\label{eq.12}
\end{equation}
Both $ R\left(\alpha^{\ast} \right) $ and $ \alpha^{\ast} $ are independent of $ Q_{1} $ and  then of $  S_{1}$.

\textit{Converse part of Costa theorem:} From (\ref{eq.4}) we can also obtain an upper bound for the channel capacity. We have:
\setlength{\arraycolsep}{0.0em}
\begin{eqnarray}
I\left(U;Y \right) -I\left(U;S_{1} \right)&{}={}&-H\left(U \mid Y \right) + H\left(U \mid S_{1} \right)\label{eq.13} \\
&{}\leq{}&-H\left(U\mid Y,S_{1} \right) +H\left(U\mid S_{1} \right)\label{eq.14} \\
&{}={}&I\left(U;Y\mid S_{1} \right) \label{eq.15}\\
&{}\leq{}&I\left( X;Y\mid S_{1} \right)\label{eq.16} 
\end{eqnarray}
\setlength{\arraycolsep}{5pt}
where inequality (\ref{eq.14}) follows from the fact that conditioning reduces the entropy and (\ref{eq.16}) follows from Markov chain $ U \rightarrow XS_{1}\rightarrow Y $ which is correct for all distributions $ p\left(y,x,u,s_{1}\right) $ in the form of (\ref{eq.5}), including the distributions in the set $\mathcal{P}_{C}$. Hence we can write:
\setlength{\arraycolsep}{0.0em}
\begin{eqnarray}
C_{Costa}&{}={}&\max_{p\left( u,x \mid s_{1}\right) }\left[I\left(U;Y \right) -I\left(U;S_{1} \right)  \right]\label{eq.17}\\
&{}\leq{}&\max_{p\left( x \mid s_{1}\right)}\left[I\left( X;Y\mid S_{1} \right)  \right]\label{eq.18}\\ 
&{}={}&\max_{p\left(x\mid s_{1}\right)}\left[H\left(Y\mid S_{1}\right) -H\left(Y\mid X,S_{1} \right)  \right]\label{eq.19}\\
&{}={}&\max_{p\left(x\mid s_{1}\right)}\left[H\left(X+Z\mid S_{1}\right) -H\left(Z \mid X,S_{1}\right)  \right]\label{eq.20}\\
&{}\leq{}&\max_{p\left(x\mid s_{1}\right)}\left[H\left(X+Z\right) -H\left(Z \right)  \right]\label{eq.21}\\
&{}={}&\frac{1}{2}\log\left( 1+\frac{P}{N}\right), \label{eq.22}
\end{eqnarray}
\setlength{\arraycolsep}{5pt}
where the inequality (\ref{eq.21}) is due to the fact that conditioning reduces the entropy. The maximum in  (\ref{eq.21}) is obtained when $X$ and $Z$ are jointly Gaussian with $ E\left\lbrace X^{2} \right\rbrace = P $ because when the variance is limited, Gaussian distribution maximizes the entropy. From (\ref{eq.11}) and (\ref{eq.22}) it is seen that the lower and the upper bounds of the capacity coincide, and therefore the channel capacity is equal to $ \frac{1}{2}\log\left( 1+\frac{P}{N}\right)  $. It is also concluded that for the channel described in C.1-C.3, the optimum condition which leads to the capacity is when  $ X\sim \mathcal{N}\left(0,P\right) $ and independent of $S_{1}$.\qed

\quad We can explain the Costa theorem more, as follows: Let consider $Y=X+S_{1}+S^{'}_{1}+Z$ with  independent Gaussian interference $S_{1}$ with power $Q_{1}$, $S^{'}_{1}$ with power $Q^{'}_{1}$ and $Z$ with power $N$. If the transmitter knows nothing about this interference, then we take $U=X$ and $C=\frac{1}{2}\log\left( 1+\frac{P}{N+Q_{1}+Q^{'}_{1}}\right) $. If $S_{1}$ is known at the transmitter, then we take $U=X+ \alpha S_{1}$ and we have $C=\frac{1}{2}\log\left( 1+\frac{P}{N+Q^{'}_{1}}\right) $ and if $S_{1}$ and $S^{'}_{1}$ are both known at the transmitter, then $U=X+\alpha S_{1}+\beta S^{'}_{1}$ and $C=\frac{1}{2}\log\left( 1+\frac{P}{N}\right) $.

\section{Capacity Theorem For The Gaussian Channel with Two-sided Input-Noise Dependent Side Information}\label{sec.capacity theorem} 
In this section we introduce a Gaussian channel in the presence of two-sided state information correlated to the channel input and noise. Then we present our capacity theorem for this Gaussian channel. The theorem obtains the capacity of channel in the case the channel input $X$, the side information $(S_{1},S_{2})$ and the channel noise $Z$, form the Markov chain $X \rightarrow \left (S_{1},S_{2}\right )\rightarrow Z$. With our new description of the ‍‍``re-cognition" of the transmitter on the channel noise, the probabilistic dependency between the side information $(S_{1},S_{2})$ and the channel noise $Z$, determines the cognition on the channel noise that the side information carries to the transmitter. Therefore, this Markov chain states that the transmitter  acquires all its knowledge on the channel noise just from the side information $(S_{1},S_{2})$, which is practically meaningful and acceptable in our scenario. To prove the theorem, we obtain a general achievable rate for the channel capacity (lemma 1) and then a general upper bound for the channel capacity in mentioned case (lemma 2) and show the coincidence of these lower and upper bounds.\\

\subsection{Definition of the Channel}\label{subsec.definition}
\begin{figure}[!t]
\centering
\includegraphics[width=2in]{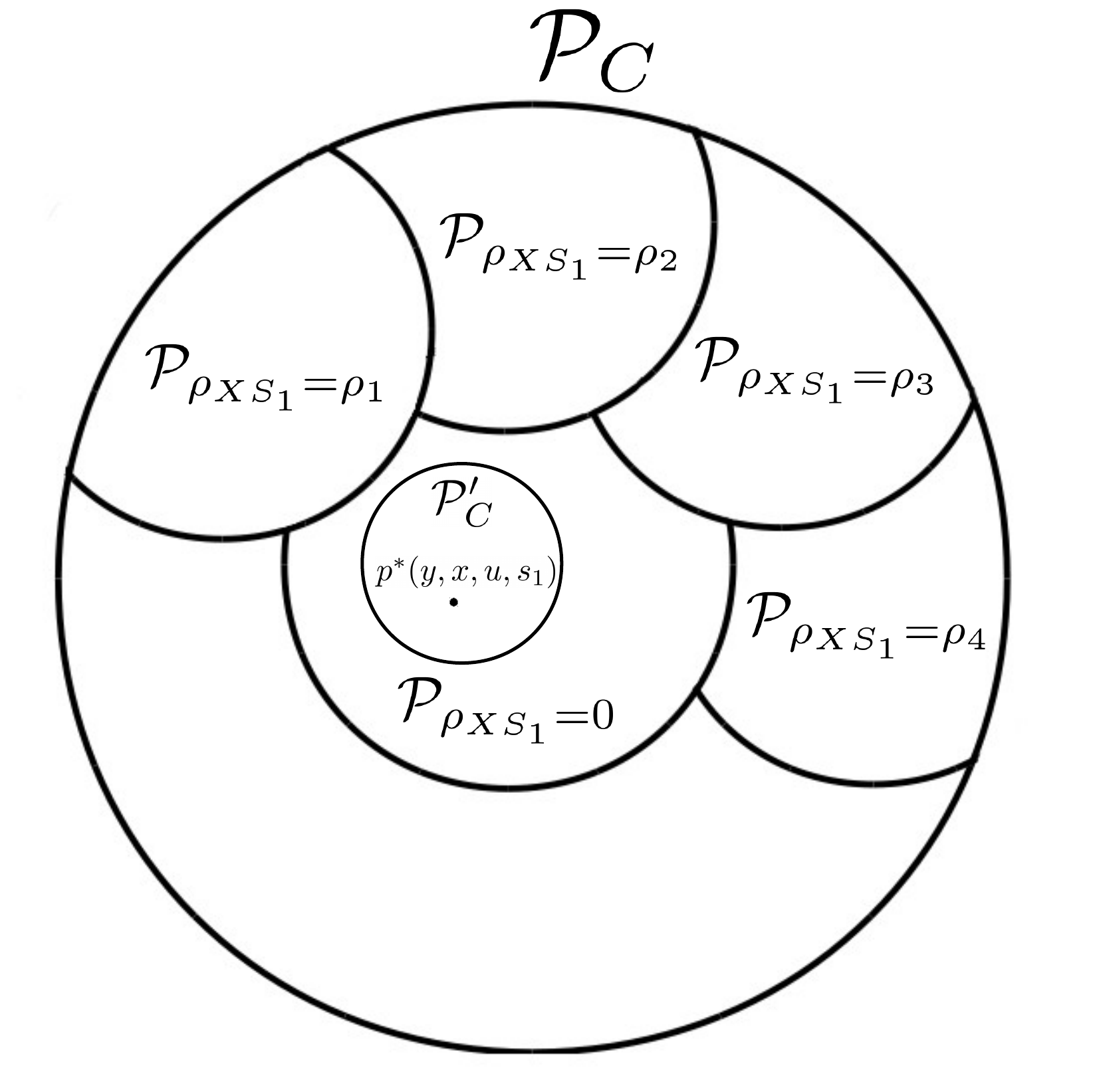}
\caption{Partitioning $\mathcal{P}_{C}$ into $\mathcal{P}_{\rho_{XS_{1}}}$'s. $p^{\ast}(y,x,u,s_{1})$ is the optimum distribution for the Costa channel.}
\label{figure7}
\end{figure}

As mentioned before, in a Gaussian channel with side information known at the transmitter defined by the set $\mathcal{P}_{C}$  with properties C.1-C.3 (Costa channel), no restriction  is imposed upon the correlation between the channel input $X$ and the side information $S_{1}$. As mentioned in section I, the capacity  $\frac{1}{2}\log \left (1+\frac{P}{N}\right )$ is only valid for channels in which $X$ and $S_{1}$ has the chance to be independent. Specifically the maximum rate is achieved when $X$ and $S_{1}$ are independent. Let $\mathcal{P}_{C}$ is partitioned into subsets $\mathcal{P}_{\rho_{XS_{1}}}$ including the distributions $ p\left(y,x,u,s_{1}\right) $ for which the correlation coefficient between $X$ and $S_{1}$ is equal to $\rho_{XS_{1}}$ as depicted in Fig. \ref{figure7}. It is obvious that $\mathcal{P}'_{C}$ (the set of distributions with properties C.1-C.5) is a subset of $\mathcal{P}_{\rho_{XS_{1}}=0}$ and therefore the optimum  distribution leading to the capacity of the Costa channel does not belong to other partitions. We can therefore claim that the Costa theorem is not valid for channels defined with random variables $(Y,X,U,S_{1})\sim p(y,x,u,s_{1})$ in partition $\mathcal{P}_{\rho_{XS_{1}}}$ with $\rho_{XS_{1}}\neq 0$. \\

Consider the Gaussian channel depicted in Fig. \ref{figure2}. The side information at the transmitter $S_{1}$ and at the receiver $S_{2}$ is considered as additive interference at the receiver. From the above discussion, providing our mentioned motivations in section I, our channel has three differences with Costa's one as follows:

1) In our channel, a specified correlation coefficient $\rho_{XS_{1}}$ between $X$ and $S_{1}$, exists.

2) To investigate the effect of the side information known at the receiver, we suppose that in our channel there exists a Gaussian side information $S_{2}$ known non-causally at the receiver which is correlated to both $X$ and $S_{1}$.

3) We allow the channel input $X$ and the side information $S_{1}$ and $S_{2}$ to be correlated to the channel noise $Z$.\\
 
\textbf{Remark:}\quad It is important to note that, as we prove in lemma 3 in the Appendix C, assuming the input random variable $X$ correlated to $S_{1}$ and $S_{2}$ with specified correlation coefficients, does not impose any restriction on $X$'s own distribution and the distribution of $X$ is still free to choose. \\

Considering the above differences, our channel is defined by the following properties GC.1-GC.4 (GC for General version of Costa) below:

\paragraph*{\textbf{GC.1}}
\quad $\left( S_{1}^{n},S_{2}^{n}\right)   $ are i.i.d.  sequences with zero mean and jointly Gaussian distributions with power $\sigma_{S_{1}}^{2}=Q_{1}$ and $\sigma_{S_{2}}^{2}=Q_{2}$ respectively (so we have $ S_{1}\sim \mathcal{N}\left(0,Q_{1} \right)  $ and $ S_{2}\sim \mathcal{N}\left(0,Q_{2} \right)  $).

\paragraph*{\textbf{GC.2}}
The output sequence  $ Y^{n}=X^{n}+S_{1}^{n}+S_{2}^{n}+Z^{n} $, where $ Z^{n} $ is the sequence of white Gaussian noise with zero mean and power $ N $ $ \big ( Z\sim \mathcal{N}\left( 0,N\right)    \big)$. The sequences $S_{1}^{n}$ and $S_{2}^{n}$ are non-causally known at the transmitter and at the receiver respectively.

\paragraph*{\textbf{GC.3}}
Random variables $ \left(X,S_{1},S_{2},Z \right)  $ have the covariance matrix $ \boldsymbol{K} $:
\begin{eqnarray}
\boldsymbol{K}{}={}E\left\lbrace   \begin{bmatrix}
X^{2} & XS_{1}& XS_{2} & XZ\\
XS_{1} & S_{1}^{2} & S_{1}S_{2} & S_{1}Z\\
XS_{2} &  S_{1}S_{2} & S_{2}^{2}& S_{2}Z \\
XZ & S_{1}Z & S_{2}Z & Z^{2}
\end{bmatrix}\right\rbrace  \label{eq.23}\\
{}={}\begin{bmatrix}
\sigma_{X}^{2}&\sigma_{X}\sigma_{S_{1}}\rho_{XS_{1}}&\sigma_{X}\sigma_{S_{2}}\rho_{XS_{2}}&\sigma_{X}\sigma_{Z}\rho_{XZ}\\
\sigma_{X}\sigma_{S_{1}}\rho_{XS_{1}}&\sigma_{S_{1}}^{2}&\sigma_{S_{1}}\sigma_{S_{2}}\rho_{S_{1}S_{2}}&\sigma_{S_{1}}\sigma_{Z}\rho_{S_{1}Z}\\
\sigma_{X}\sigma_{S_{2}}\rho_{XS_{2}}&\sigma_{S_{1}}\sigma_{S_{2}}\rho_{S_{1}S_{2}}&\sigma_{S_{2}}^{2}&\sigma_{S_{2}}\sigma_{Z}\rho_{S_{2}Z}\\
\sigma_{X}\sigma_{Z}\rho_{XZ}&\sigma_{S_{1}}\sigma_{Z}\rho_{S_{1}Z}&\sigma_{S_{2}}\sigma_{Z}\rho_{S_{2}Z}&\sigma_{Z}^{2}
\end{bmatrix}\label{eq.24}
\end{eqnarray}
and therefore, in our channel, the Gaussian noise $Z$ is not necessarily independent of the additive interference $S_{1}$ and $S_{2}$ and the input $X$.  Moreover  $ X^{n} $ is assumed to have the constraint $ \sigma_{X}^{2}\leq P  $. Except $ \sigma_{X} $, all other parameters in $ \boldsymbol{K} $ have fixed values specified for the channel and must be considered as \textit{the definition of the channel}.

\paragraph*{\textbf{GC.4}}
\quad $ \left( X,U,S_{1},S_{2}\right)$ form the Markov Chain  $ S_{2}\rightarrow S_{1}\rightarrow UX $. As mentioned earlier, this Markov chain is satisfied by all distributions $ p\left(y,x,u,s_{1},s_{2}\right)$ in the form of (\ref{eq.2}) in Cover-Chiang capacity theorem and is physically reasonable.
Since this Markov chain results in the weaker Markov chain $ S_{2}\rightarrow S_{1}\rightarrow X $, as proved in lemma 4 in the Appendix D,  this property implies that in the covariance matrix $ \boldsymbol{K} $ in (\ref{eq.24}) we have:
\begin{equation}
\rho_{XS_{2}}=\rho_{XS_{1}}\rho_{S_{1}S_{2}} \label{eq.43+1}
\end{equation} 

It is readily seen that all distributions $ p\left(y,x,u,s_{1},s_{2}\right)$ having the properties GC.1-GC.4 are in the form of (\ref{eq.2}). Therefore we can apply the extended version of Cover-Chiang theorem for random variables with continuous alphabets to our channel. We denote the set of all these distributions $ p\left(y,x,u,s_{1},s_{2}\right)$ with $\mathcal{P}_{\rho_{XS_{1}}}$ (again).\\

\textbf{Remark:}\quad In the absence of $S_{2}$ and when $Z$ is independent of $\left (X,S_{1}\right )$, we can compare the capacity of our channel with the Costa channel and write:
\begin{equation}
C_{Costa}=\max_{S_{2}=0 ,\rho_{XS_{1}}}C_{1}.\label{eq.34+1} 
\end{equation}\\
where $C_{1}$ denotes the capacity of our channel when $Z$ is independent of $\left (X,S_{1},S_{2}\right )$. Note that in this case and when $S_{2}=0$, we have $\mathcal{P}_{C}=\bigcup_{\rho_{XS_{1}}}\mathcal{P}_{\rho_{XS_{1}}}$ and therefore, looking for the maximum rate in $\mathcal{P}_{C}$ leads to the maximum rate among $\mathcal{P}_{\rho_{XS_{1}}}$'s.\\

We will show that the optimum distribution resulting in maximum transmission rate, is obtained when $(X,S_{1},S_{2})$ are jointly Gaussian  and the auxiliary random variable $U$ is a linear combination of $X$ and $S_{1}$. We denote the set of distributions $ p^{\ast}\left(y,x,u,s_{1},s_{2}\right)$ having properties GC.5 and GC.6 below as well as properties GC.1-GC.4, with $\mathcal{P}_{\rho_{XS_{1}}}^{\ast}$:

\paragraph*{GC.5}
 \quad The random variables $ \left( X,S_{1},S_{2}\right)$ are jointly Gaussian distributed and $X$  has zero mean and the maximum power $ P $ i.e. $ X\sim \mathcal{N}\left(0,P \right)  $.\\

\paragraph*{GC.6}
\quad As in the Costa theorem:
\begin{equation}
U=\alpha S_{1}+X.\label{eq.36}
\end{equation}
where $X$ and $S_{1}$ are now correlated.

It is clear that the set $\mathcal{P}_{\rho_{XS_{1}}}^{\ast}$ (described in GC.1-GC.6) and their marginal and conditional distributions are subsets of corresponding $\mathcal{P}_{\rho_{XS_{1}}}$'s (described in GC.1-GC.4).\\

As the final part of this subsection we introduce some definitions required for our capacity theorem:

Suppose $\widehat{\boldsymbol{K}}$ is the covariance matrix for random variables $(X,S_{1},S_{2},Z)$ having all properties GC.1-GC.6; defining:
\setlength{\arraycolsep}{0.0em}
\begin{eqnarray}
A_{i} &=& E\left\lbrace XS_{i}\right\rbrace=\sigma_{X}\sigma_{S_{i}}\rho_{XS_{i}}\quad ,i=1,2 \label{eq.34+2}\\
L_{0}&=&E\left\lbrace XZ\right\rbrace=\sigma_{X}\sigma_{Z}\rho_{XZ}\label{eq.34+3} \\
L_{i} &=& E\left\lbrace S_{i}Z\right\rbrace=\sigma_{S_{i}}\sigma_{Z}\rho_{S_{i}Z}\quad  ,i=1,2
\label{eq.34+4} \\
B &=& E\left\lbrace S_{1}S_{2}\right\rbrace =\sigma_{S_{1}}\sigma_{S_{2}}\rho_{S_{1}S_{2}} \label{eq.34+5} 
\end{eqnarray}
\setlength{\arraycolsep}{5pt}
we can write $\widehat{\boldsymbol{K}}$, its determinant $D$ and its minors as:
\begin{equation}
\widehat{\boldsymbol{K}}=\begin{bmatrix}
P&A_{1}&A_{2}&L_{0}\\
A_{1}&Q_{1}&B&L_{1}\\
A_{2}&B&Q_{2}&L_{2}  \\
L_{0}&L_{1}&L_{2}&N\\
\end{bmatrix}.\label{eq.35}
\end{equation}\\
\setlength{\arraycolsep}{0.0em}
\begin{eqnarray}
D&\triangleq & \begin{vmatrix}
P&\quad A_{1}&\quad A_{2}&\quad L_{0}\\
A_{1}&\quad Q_{1}&\quad B&\quad L_{1}\\
A_{2}&\quad B&\quad Q_{2}&\quad L_{2}  \\
L_{0}&\quad L_{1}&\quad L_{2}&\quad N\\
\end{vmatrix} \label{eq.37+1} 
\end{eqnarray}
\setlength{\arraycolsep}{5pt}

\setlength{\arraycolsep}{0.0em}

\begin{equation}
\left .
\begin{array}{rclcrclcrcl}

d_{P}&\triangleq & \begin{vmatrix}
Q_{1}&\quad B&\quad L_{1}\\
B&\quad Q_{2}&\quad L_{2}\\
L_{1}&\quad L_{2}&\quad N
\end{vmatrix}
& \quad  \text{,} &  \quad  d_{Q_{1}}&\triangleq & \begin{vmatrix}
P&\quad A_{2}&\quad L_{0}\\
A_{2}&\quad Q_{2}&\quad L_{2}\\
L_{0}&\quad L_{2}&\quad N
\end{vmatrix}
& \quad  \text{,}  &  \quad  d_{A_{1}}&\triangleq & \begin{vmatrix}
A_{1}&\quad B&\quad L_{1}\\
A_{2}&\quad Q_{2}&\quad L_{2}\\ 
L_{0}&\quad L_{2}&\quad N
\end{vmatrix}\\

 \\
 
d_{L_{0}}&\triangleq & \begin{vmatrix}
A_{1}&\quad Q_{1}&\quad B\\
A_{2}&\quad B&\quad Q_{2}\\
L_{0}&\quad L_{1}&\quad L_{2}
\end{vmatrix}
&  \quad \text{,}  \quad & d_{L_{1}}&\triangleq & \begin{vmatrix}
P&\quad A_{1}&\quad A_{2}\\
A_{2}&\quad B&\quad Q_{2}\\
L_{0}&\quad L_{1}&\quad L_{2}
\end{vmatrix}
& \quad  \text{,}  & \quad d_{N}&\triangleq & \begin{vmatrix}
P&\quad A_{1}&\quad A_{2}\\
A_{1}&\quad Q_{1}&\quad B\\
A_{2}&\quad B&\quad Q_{2}
\end{vmatrix} \\

 \\

d_{Q_{1}N}&\triangleq &  \begin{vmatrix}
P&\quad A_{2}\\
A_{2}&\quad Q_{2}
\end{vmatrix}
&  \quad \text{,}  \quad & d_{Q_{2}N}&\triangleq & \begin{vmatrix}
P&\quad A_{1}\\
A_{1}&\quad Q_{1}
\end{vmatrix}
& \quad  \text{,}  & \quad d_{PN}&\triangleq &  \begin{vmatrix}
Q_{1}&\quad B\\
B&\quad Q_{2}
\end{vmatrix} \\

 \\

d_{PQ_{1}}&\triangleq &  \begin{vmatrix}
Q_{2}&\quad L_{2}\\
L_{2}&\quad N
\end{vmatrix}
&  \quad \text{,}  \quad & d_{L_{0}L_{1}}&\triangleq &  \begin{vmatrix}
A_{1}&\quad A_{2}\\
B&\quad Q_{2}
\end{vmatrix}
& \quad  \text{,}  & \quad d_{PL_{1}}&\triangleq &  \begin{vmatrix}
B&\quad Q_{2}\\
L_{1}&\quad L_{2}
\end{vmatrix} \\

 \\

d_{Q_{1}L_{0}}&\triangleq &  \begin{vmatrix}
A_{2}&\quad Q_{2}\\
L_{0}&\quad L_{2}
\end{vmatrix}
&  \quad   \quad & 
& \quad   & \quad

\end{array}
\quad \quad \quad \quad \right\} \label{eq.300}
\end{equation}
\setlength{\arraycolsep}{5pt}

\subsection{The Capacity of the Channel}\label{subsec.theorem}
\subsubsection*{Theorem} The Gaussian channel defined by properties GC.1-GC.4, when the channel input $X$, the side information $(S_{1},S_{2})$ and the channel noise $Z$ form the Markov chain $X \rightarrow \left (S_{1},S_{2}\right )\rightarrow Z$, has the capacity:\\
\begin{equation}
C=\dfrac{1}{2}\log\left( 1+\dfrac{P}{N}\dfrac{\left (1-\rho_{XS_{1}}^{2}\right ) \left (1-\rho_{S_{1}S_{2}}^{2}\right )}{\quad d_{P}^{\mathcal{N}}} \right ),\label{eq.95+1}
\end{equation}
where
\begin{eqnarray}
d_{P}^{\mathcal{N}}&=&\begin{vmatrix}
1&\rho_{S_{1}S_{2}}&\rho_{S_{1}Z}\\
\rho_{S_{1}S_{2}}&1&\rho_{S_{2}Z}\\
\rho_{S_{1}Z}&\rho_{S_{2}Z}&1\\
\end{vmatrix} \nonumber \\
&=&1+2\rho_{S_{1}S_{2}}\rho_{S_{1}Z}\rho_{S_{2}Z}-\rho_{S_{1}S_{2}}^{2}-\rho_{S_{1}Z}^{2}-\rho_{S_{2}Z}^{2}. \label{eq.96xx}
\end{eqnarray}
\paragraph*{Proof of Theorem} To prove the theorem, first, we prove a general achievable rate for the channel in lemma 1. Then in lemma 2, we obtain  an upper bound for the channel in the case the transmitter acquires all its knowledge on the channel noise $Z$ from the side information $\left(S_{1},S_{2}\right ) $, i.e, we have the Markov chain $X \rightarrow \left (S_{1},S_{2}\right )\rightarrow Z$. Then we show the coincidence of this upper bound with the lower bound of the capacity.

We note that the Markov chain $X \rightarrow \left (S_{1},S_{2}\right )\rightarrow Z$ and the Markov chain $X\rightarrow S_{1}\rightarrow S_{2}$ from GC4, imply the weaker Markov chain $X\rightarrow S_{1} \rightarrow Z$. And since $S_{1}$ and $Z$ are Gaussian, as we prove in lemma 4 in the Appendix D, the recent Markov chain implies that 
\begin{equation}
 \rho_{XZ}=\rho_{XS_{1}}\rho_{S_{1}Z}. \label{eq.300x} 
 \end{equation} 

\subsubsection*{Lemma 1. A General Lower Bound for the Capacity of the Channel}
The capacity of the Gaussian channel defined with properties GC.1-GC.4 has the lower bound:
\begin{equation}
R_{G}=\dfrac{1}{2}\log\left( 1+\dfrac{\left [\sigma_{X}\left (1-\rho_{XS_{1}}^{2}\right )-\sigma_{Z}\left (\rho_{XS_{1}}\rho_{S_{1}Z}-\rho_{XZ}\right )\right ]^{2}\left (1-\rho_{S_{1}S_{2}}^{2}\right )}{\sigma_{Z}^{2}\left (\left (1-\rho_{XS_{1}}^{2}\right )d_{P}^{\mathcal{N}}-\left (\rho_{XS_{1}}\rho_{S_{1}Z}-\rho_{XZ}\right )^{2}\left (1-\rho_{S_{1}S_{2}}^{2}\right )\right )} \right )\label{eq.45}
\end{equation}
where $d_{P}^{\mathcal{N}}$ is defined in (\ref{eq.96xx}).
\paragraph*{Proof} Appendix A contains the proof.\\

\subsubsection*{Lemma 2. Upper Bound for the Capacity of the Channel}
The capacity of the Gaussian channel defined by properties GC.1-GC.4, when the channel input $X$, the side information $(S_{1},S_{2})$ and the channel noise $Z$ form the Markov chain $X \rightarrow \left (S_{1},S_{2}\right )\rightarrow Z$, has the upper bound  $C$ in (\ref{eq.95+1}).  
\paragraph*{Proof} Appendix B contains the proof.\\

For completing the proof of the theorem, it is enough to compute the lower bound of the channel (\ref{eq.45}), when we have the Markov chain $X \rightarrow \left (S_{1},S_{2}\right )\rightarrow Z$. Applying the equation (\ref{eq.300x}) to equation (\ref{eq.45}), shows the coincidence of the upper and the lower bounds of the capacity of the channel in this case and the proof is completed.  \qed \\
\textbf{Remark 1:} It can be shown that for variables $S_{1}$, and $S_{2}$ and $Z$ with properties GC.1 and GC.4: 
\begin{equation}
I\left (S_{1}S_{2};Z \right )=\dfrac{1}{2} \log \left ( \dfrac{1-\rho_{S_{1}S_{2}}^{2}}{d_{P}^{\mathcal{N}}} \right ) \label{eq.95+I} 
\end{equation}
and so the channel capacity (\ref{eq.95+1}) can be written as:
\begin{equation}
C=\dfrac{1}{2}\log\left( 1+\dfrac{P}{N} \left (1-\rho_{XS_{1}}^{2}\right ) \exp \left (  2I\left (S_{1}S_{2};Z \right ) \right ) \right ),\label{eq.95+C}
\end{equation}
that is an increasing function of $I\left (S_{1}S_{2};Z \right )$.\\
\textbf{Remark 2:} The transmission rate $C$ in (\ref{eq.95+1}) can be reached by encoding and decoding schema represented in \cite{coverchiang} modified for continuous Gaussian distributions.

\section{Interpretations and Numerical Results of the Capacity Theorem}
In previous section, the capacity of the Gaussian channel with two-sided information correlated to the channel input and noise, has been obtained. The capacity theorem is general except that the Markov chain $X \rightarrow \left (S_{1},S_{2}\right )\rightarrow Z$ must be satisfied. In this section we present some corollaries of the capacity theorem. First, we examine the effect of the correlation between the side information and the channel input on the channel capacity.  Second, we try to exemplify our new description of the concept of ``cognition" of a communicating object (here, transmitter and or receiver) on some features of channel (here, channel noise), by our capacity theorem.
\subsection{The Effect of the Correlation between the Side Information and the Channel Input on the Capacity:} If we assume that the channel noise $Z$ is independent of $\left (X,S_{1},S_{2} \right )$, from (\ref{eq.95+1}), the capacity of the channel is:
\begin{equation}
 C_{1}=\frac{1}{2} \log \left (1+\dfrac{P}{N} \left (1-\rho_{XS_{1}}^{2} \right )\right ) \label{eq.150}
 \end{equation} 
 
\textit{Corollary 1:} From (\ref{eq.34+1}), $C_{1}$ is reduced to the Costa capacity by maximizing it with $\rho_{XS_{1}}=0$.

\textit{Corollary 2:} It is seen that in the case the side information $S_{2}$ is independent of the channel noise $Z$, the capacity of the channel is equal to the capacity when there is no interference $S_{2}$. In other words, in this case, the receiver can subtract the known $S_{2}^{n}$ from the received $Y^{n}$ without losing any worthy information. 

\textit{Corollary 3:} The correlation between $X$ and $S_{1}$ decreases the capacity of the channel. It can be explained as follows:  by looking at $Y=X+S_{1}+Z$ in our dirty paper like coding, mitigating the input-dependent interference effect, also mitigates the input power impact on the channel capacity as this fact is seen in (\ref{eq.150}) as $\sigma_{X}^{2}\left (1-\rho_{XS_{1}}^{2} \right )$. 

As an extreme and interesting case, when $S_{1}=X$ (then $\rho_{XS_{1}}=1 $), according to the usual Gaussian coding, the capacity seems to be $\frac{1}{2}\log\left ( 1+\frac{4P}{N}\right )$, which is the capacity when $2X$ is transmitted and $Y=2X+Z$ is received. But as our theorem shows, the capacity paradoxically is zero. Because the receiver based on his information ought to decode according to the dirty paper like coding. In DP like coding, with given known sequence $S_{1,0}^{n}$, we must find an auxiliary sequence $U^{n}$ like $U_{0}^{n}$ jointly typical with $S_{1,0}^{n}$ \cite{costa}. Jointly typicality of $(U_{0}^{n},S_{1,0}^{n})$ is equivalent to:
\begin{equation}
\left | \left (U_{0}^{n}-\alpha^{\ast}S_{1,0}^{n}\right )^{T}S_{1,0}^{n} \right | \leq \delta \quad , \quad \text{$ \delta $ small} \label{eq.151} 
\end{equation}
where $.^{T}$ denotes the transpose operation and $\alpha^{\ast}$ is computed according to (\ref{eq.48}). If $X=S_{1}$, there exists no such $U_{0}^{n}$: since $X_{0}^{n}=U_{0}^{n}-\alpha^{\ast}S_{1,0}^{n}=S_{1,0}^{n}$,  we have
\begin{equation}
\left | \left (U_{0}^{n}-\alpha^{\ast}S_{1,0}^{n}\right )^{T}S_{1,0}^{n} \right |= ||S_{1,0}^{n}||^{2}
\end{equation}
where $||S_{1,0}^{n}||$ is the norm of the given known sequence $S_{1,0}^{n}$ and therefore (\ref{eq.151}) can not be true. In other words, in this case, encoding error  occurs. 
 
\begin{figure}[!t]
\centering
\includegraphics[width=3.5in]{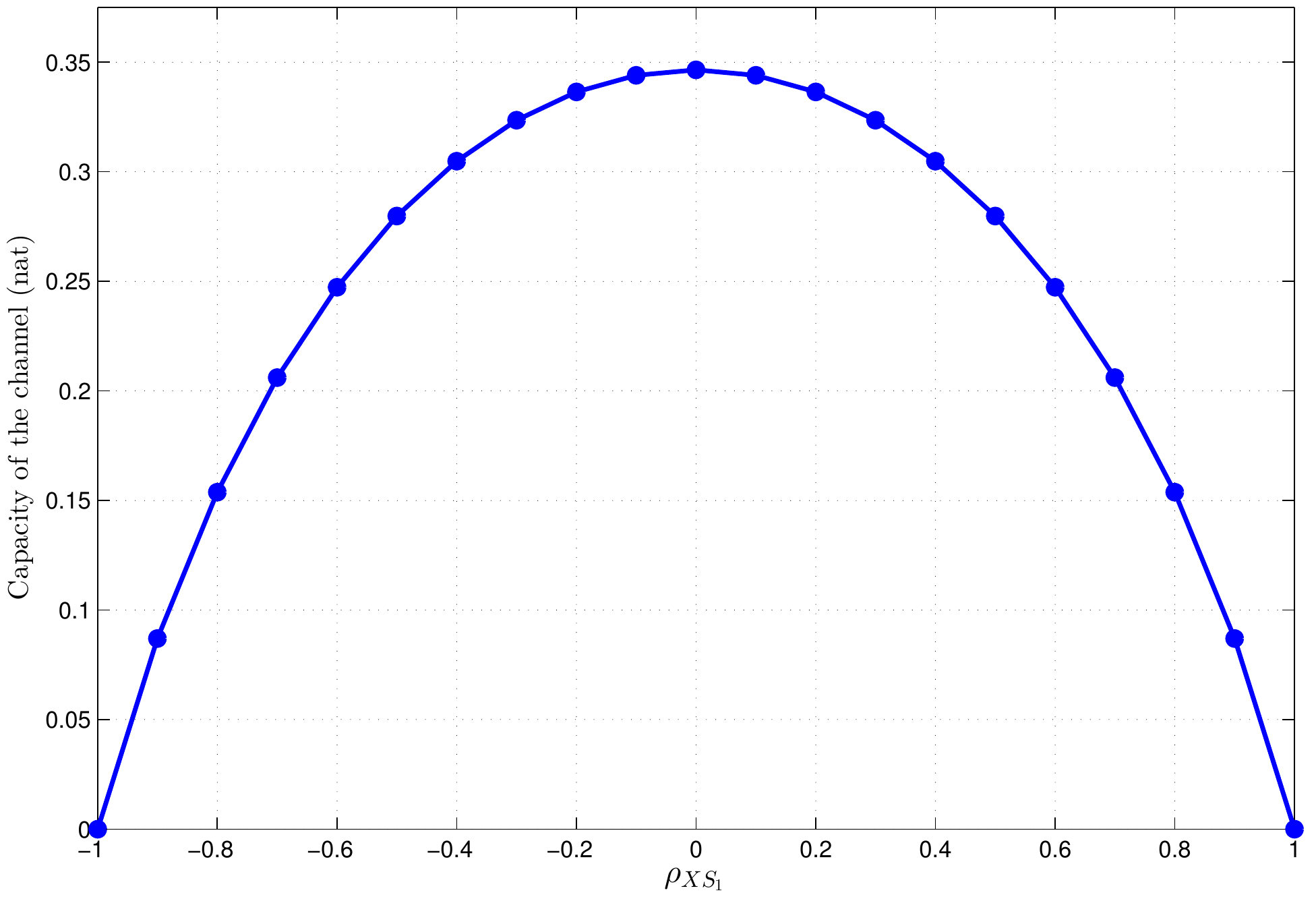}
\caption{Capacity of the channel with respect to $\rho_{XS_{1}}$ when $S_{2}=0$ and $\dfrac{P}{N}=1$.}
\label{capacity1}
\end{figure}

\begin{figure}[!t]
\centering
\includegraphics[width=3.5in]{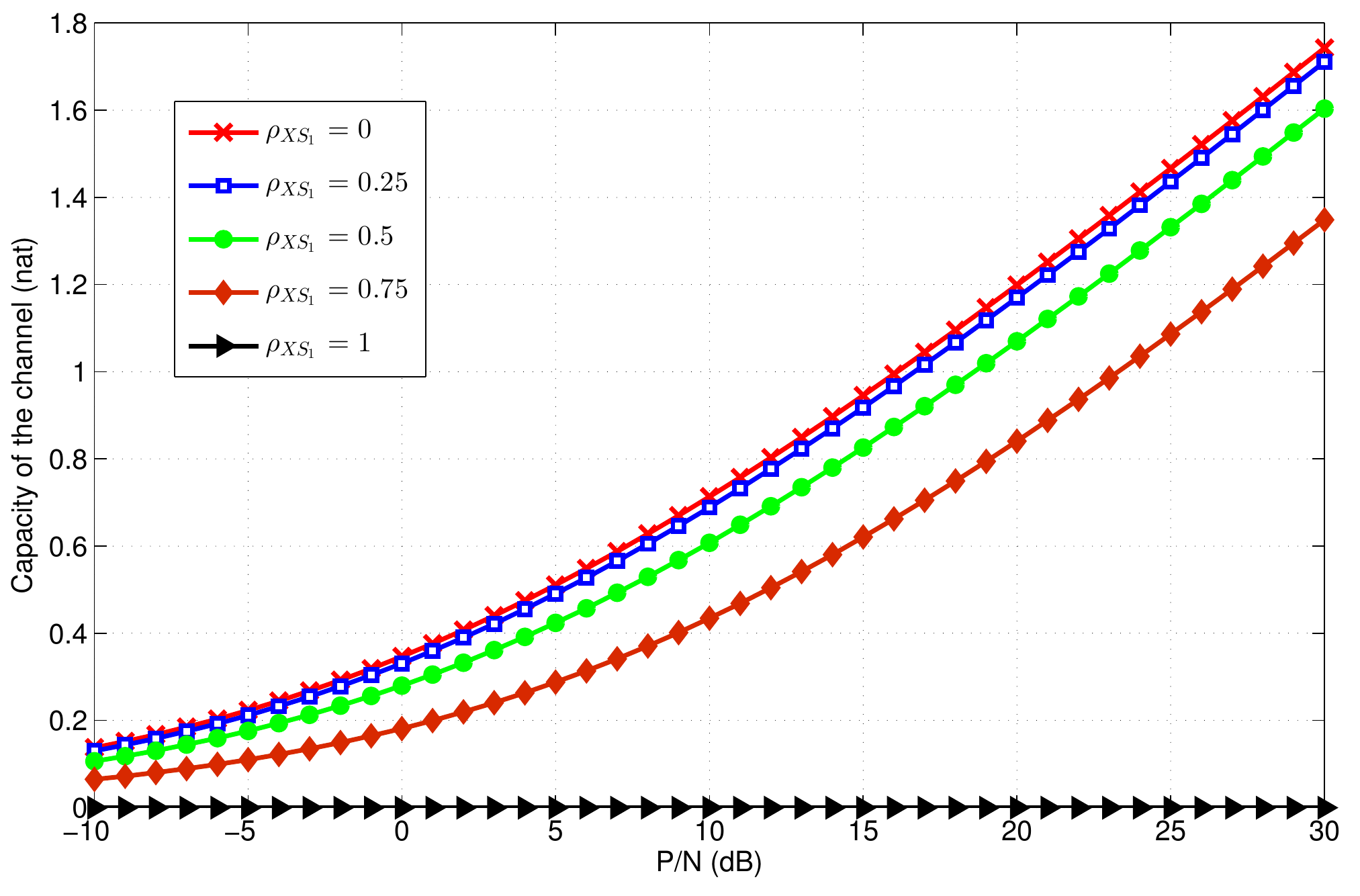}
\caption{Capacity of the channel with respect to the $SNR$ when $S_{2}=0$.}
\label{capacity2}
\end{figure}

Fig. \ref{capacity1} shows the variation of the capacity $C_{1}$ with respect to $\rho_{XS_{1}}$ when $\frac{P}{N}=1$. It is seen that when the correlation between the channel input and the side information known at the transmitter increases, the channel capacity decreases. The maximum capacity is gained when $\rho_{XS_{1}}=0$, that is Costa's capacity. Fig. \ref{capacity2} shows the capacity $C_{1}$ with respect to $SNR$ for five values of $\rho_{XS_{1}}$.

\subsection{Exemplification of the Re-cognition of Transmitter and Receiver on the channel Noise:}
\subsubsection{Re-cognition} ``Cognition" is an indispensable concept in communication. The assumption that an intelligent communicating object (transmitter, receiver, relay and so on) has got \textit{some} side knowledge about \textit{some} features of the communication channel, is a true and acceptable assumption. This exceeded information owned, for example, by the transmitter is described by "side information" known at the transmitter. In usual description, the side information is considered as the subject of cognition itself, for example, the interference of another transmitter in a cognitive radio channel \cite{DevroyeMitranTarokh2006}. On the other hand, the assumption that the knowledge may be incomplete or imperfect, is necessary in most  communication scenarios. Describing this incomplete cognition and corresponding information-theoretic concept, i.e., partial side information are found in the literature; for example in \cite{Gueguen2009} the imperfect known interference is partitioned to one perfect known and one unknown parts; and in \cite{zaidi2006} partial side information is considered as a disturbed version of the subject variable by noise.

We try here to present an alternative description for the concept of ``cognition" in communication by the concept of side information. The essential property of this description is the \textit{separation} of the \textit{subject} of knowledge $K$ (for example interference, channel noise, fading coefficients and so on) from the side information $S$ that \textit{carries} the knowledge for the intelligent agent (for example transmitter, receiver, relay and so on) and known by it. This point of view is compatible with what happens in reality: we always acquire our knowledge on something indirectly by knowing other things. What make it possible to extract the knowledge about $K$ from $S$ is dependency between $S$ and $K$. Each method of extraction of knowledge about $K$ from $S$ (estimation and so on), originally relies on this dependency. If $S$ is independent from $K$ then $S$ is non-informative about $K$. And it is expected that increasing the dependency between $S$ and $K$, increases the possible knowledge of $S$ about $K$. 

Avoiding confusion between this new with the usual descriptions of the cognition, we use the word \textbf{``re-cognition"} for it and  define it as follows: 

\textit{A communicating agent (transmitter, receiver, relay and so on) has ``re-cognition" on some variable $K$ if the side information $S$ known by it, has probabilistic dependency on $K$. }

\subsubsection{Exemplification} In the Gaussian channel defined and analyzed in the previous section, the side information $\left ( S_{1},S_{2} \right )$ is dependent to the channel noise and therefore the transmitter and the receiver have got re-cognition on the channel noise by $S_{1}$ and $S_{2}$ respectively. The capacity is proved with Markovity constraint $X \rightarrow \left (S_{1},S_{2}\right )\rightarrow Z$. Considering the new description of re-cognition, this Markov chain simply  means that the transmitter acquires all its re-cognition on the channel noise via the side information $(S_{1},S_{2})$, which is meaningful and acceptable.

\textit{Corollary 4:} If $S_{2}=0$, the transmitter have re-cognition on the channel noise $Z$ obtained by $S_{1}$ correlated to the noise. If there is no constraint on correlation between $X$ and $S_{1}$, $\rho_{XS_{1}}=0$ maximizes the transmission rate, as mentioned in (\ref{eq.34+1}). Therefore, from (\ref{eq.95+1}) and (\ref{eq.95+C}),  the capacity in this case is:
\begin{equation}
C=\frac{1}{2} \log \left (1+\dfrac{P}{N}\dfrac{1}{\left (1-\rho_{S_{1}Z}^{2} \right )} \right )=\frac{1}{2} \log \left (1+\left (\dfrac{P}{N} \right ) \exp \left (  2I\left (S_{1};Z \right ) \right ) \right ). \label{eq.152}
\end{equation}  
It is seen that more correlation between $S_{1}$ and $Z$ results in more re-cognition of the transmitter on the channel noise and more capacity. The capacity reaches to infinite when $\rho_{S_{1}Z}=\pm 1$ and therefore the transmitter has perfect re-cognition about the channel noise.

\begin{figure}[!t]
\centering
\includegraphics[width=3.5in]{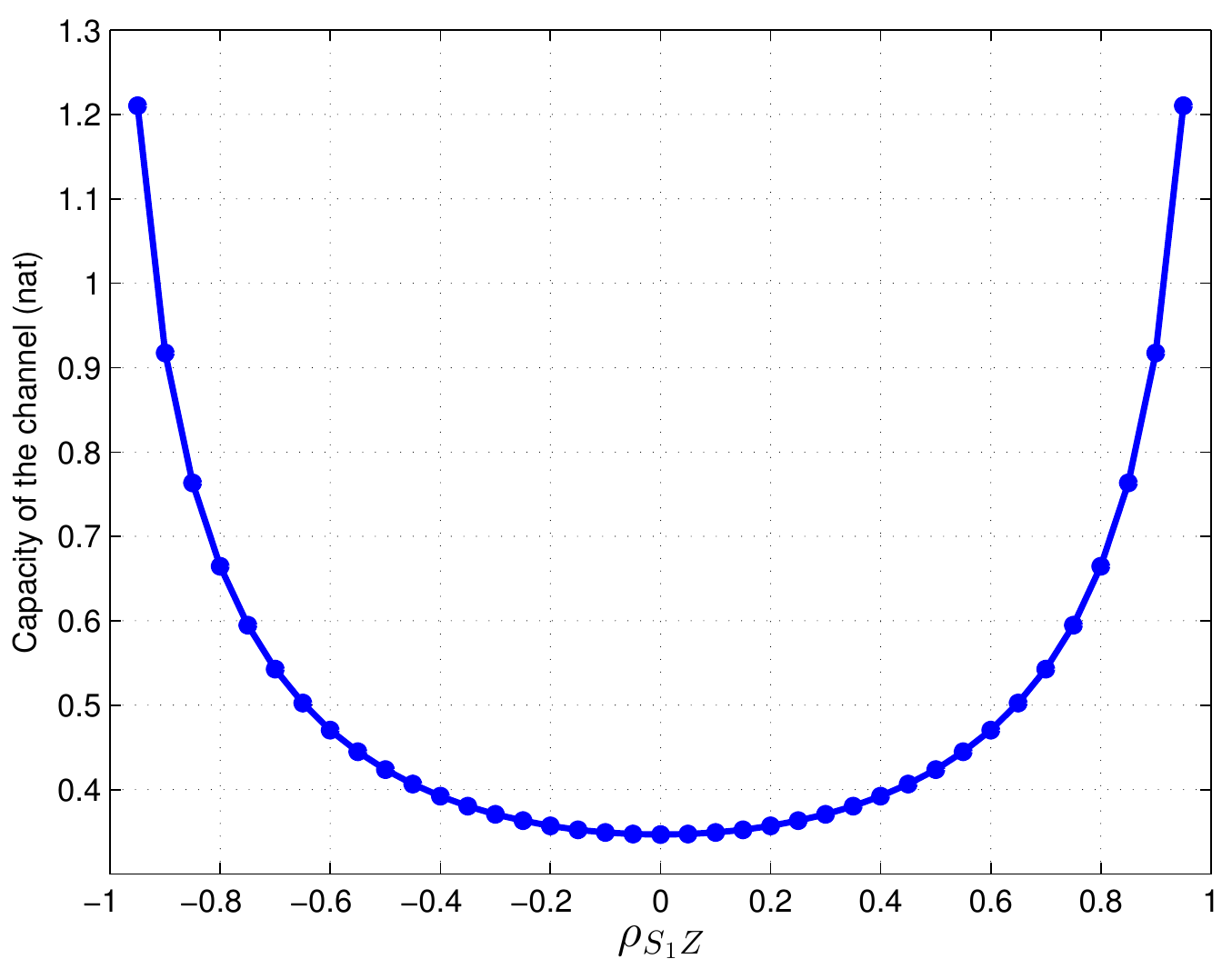}
\caption{Capacity of the channel with respect to $\rho_{S_{1}Z}$ when $\frac{P}{N}=1$.}
\label{capacity3}
\end{figure}

\begin{figure}[!t]
\centering
\includegraphics[width=3.5in]{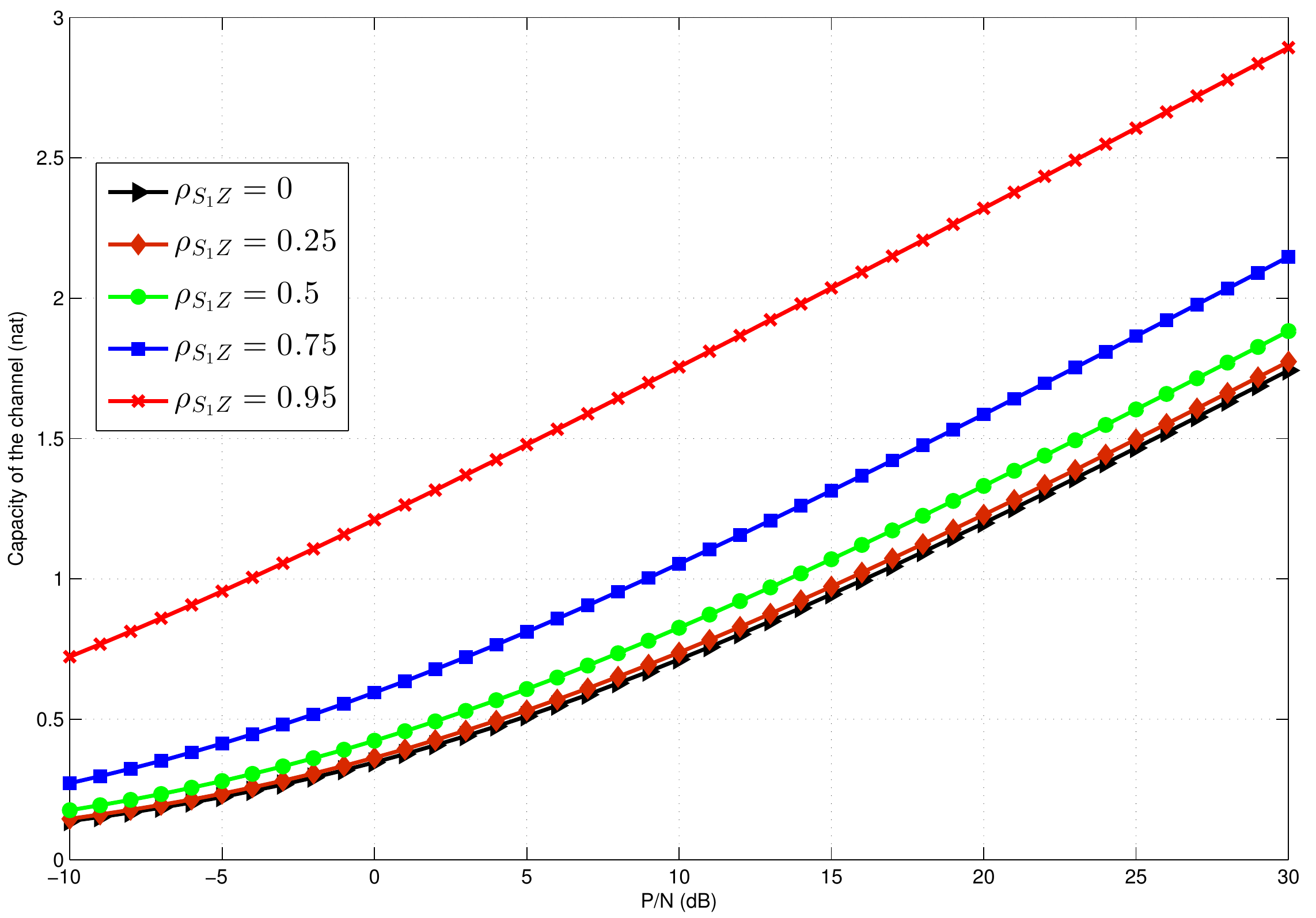}
\caption{Capacity of the channel with respect to $SNR$ for five values of $\rho_{S_{1}Z}$.}
\label{capacity4}
\end{figure}

\begin{figure}[!t]
\centering
\includegraphics[width=3.5in]{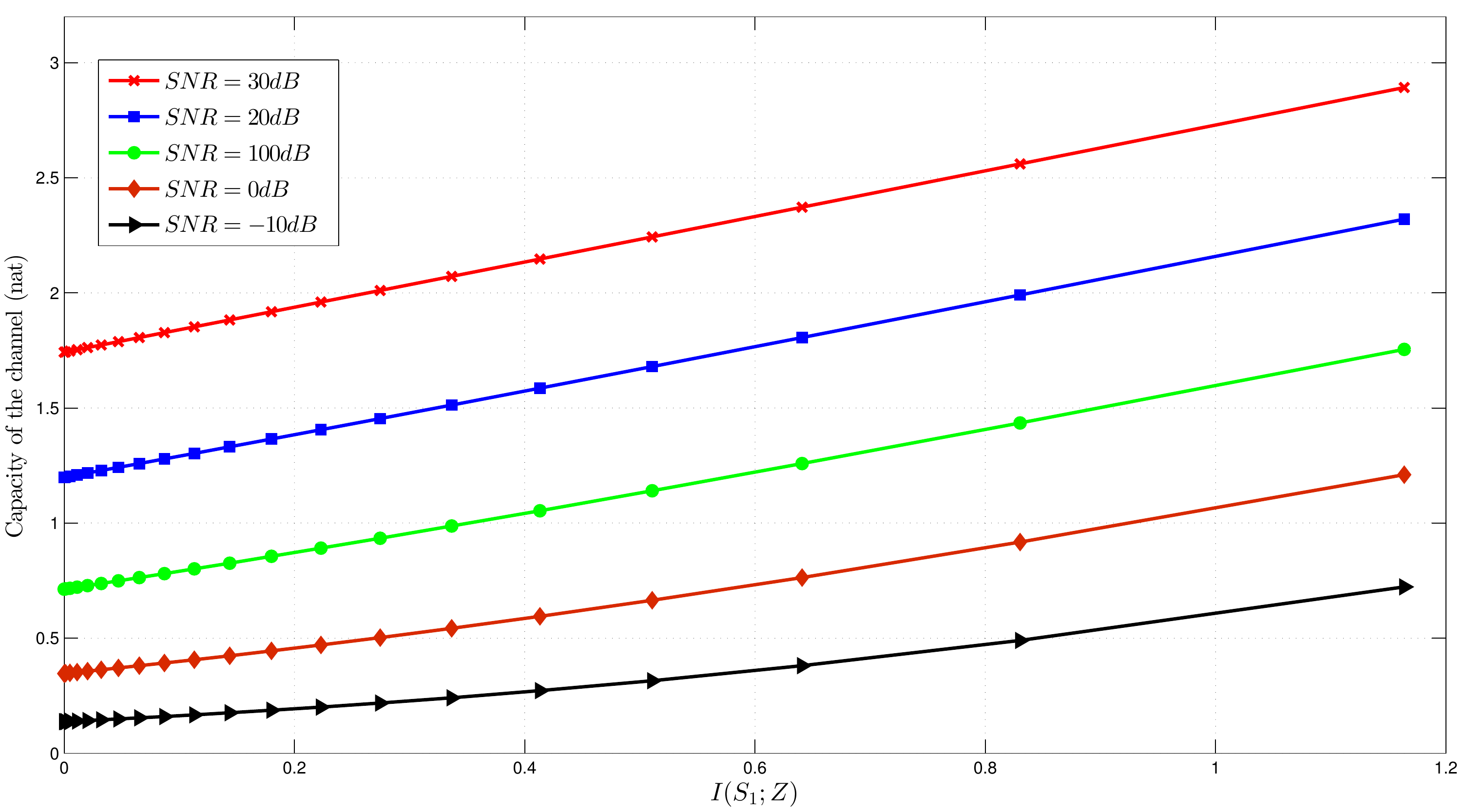}
\caption{Capacity of the channel with respect to $I\left (S_{1};Z \right )$ for five values of $SNR$.}
\label{capacity6}
\end{figure}

Fig. \ref{capacity3} illustrates the capacity of the channel with respect to $\rho_{S_{1}Z}$, the correlation coefficient between the side information $S_{1}$ and the channel noise when $\frac{P}{N}=1$. It is seen that when the correlation increases (that it means that $S_{1}$ carries more re-cognition on the channel noise to the transmitter), the capacity increases. Fig. \ref{capacity4} shows the capacity of the channel with respect to $SNR$ for five values of $\rho_{S_{1}Z}$. 
Fig. \ref{capacity6}  illustrates the capacity of the channel with respects to mutual information $I\left (S_{1};Z \right )$ for five values of $SNR$.

\textit{Corollary 5:} If $S_{1}=0$, the receiver have re-cognition on the channel noise $Z$ obtained by $S_{2}$ correlated to the noise. The capacity in this case is:
\begin{equation}
C=\frac{1}{2} \log \left (1+\dfrac{P}{N}\dfrac{1 }{ \left (1-\rho_{S_{2}Z}^{2} \right )} \right ). \label{eq.153}
\end{equation}  
It is seen that more correlation between $S_{2}$ and $Z$ results in more re-cognition of the receiver on the channel noise and more capacity. Perfect re-cognition takes place with $\rho_{S_{2}Z}=\pm 1$ and results in infinite capacity.

\textit{Corollary 6:} If $\rho_{S_{1}S_{2}}=0$, If there is no constraint on correlation between $X$ and $S_{1}$, $\rho_{XS_{1}}=0$ maximizes the transmission rate, as mentioned in (\ref{eq.34+1}). Therefore  the capacity of the channel is:
\begin{equation}
C=\frac{1}{2} \log \left (1+\dfrac{P}{N}\dfrac{1}{ \left (1-\rho_{S_{1}Z}^{2}-\rho_{S_{2}Z}^{2} \right )} \right ). \label{eq.154}
\end{equation} 
It is seen that when $\rho_{S_{1}Z}^{2}+\rho_{S_{2}Z}^{2}=1$, the capacity reaches to infinite, even if neither the transmitter nor  the receiver has perfect knowledge about the channel noise. In this case the transmitter and the receiver have their shares in re-cognition on the channel noise which leads to totally mitigating the channel noise.

\section{Conclusion}\label{conclusion} 
By fully detailed investigating the Gaussian channel in presence of two-sided input and noise dependent state information, we obtained a general achievable rate for the channel and established the capacity theorem. This capacity theorem, first demonstrate the impact of the transmitter and receiver cognition, with a new introduced interpretation on the capacity and second show the effect of the correlation between the channel input and side information available at the transmitter and at the receiver on the channel capacity. Whereas, as expected, the cognition of the transmitter and receiver increases the capacity, the correlation between the channel input and the side information known at the transmitter decreases it.

\section{Appendix}
\label{sec.VII}
\subsection*{Appendix A.}
\paragraph*{The proof of Lemma 1}
\quad Using the extension of Cover-Chiang capacity theorem given in (\ref{eq.1}) for random variables with continuous alphabets, the capacity of our channel can be written as:
\begin{equation}
C=\max_{p\left (u,x\mid s_{1}\right )}\left[ I\left (U;Y,S_{2}\right )-I\left (U;S_{1}\right )\right ]\label{eq.49}
\end{equation}
where the maximum is over all distributions $ p\left (y,x,u,s_{1},s_{2}\right ) $ in $\mathcal{P}_{\rho_{XS_{1}}}$ having properties GC.1-GC.4. Since $\mathcal{P}_{\rho_{XS_{1}}}^{\ast}\subseteq \mathcal{P}_{\rho_{XS_{1}}}$ we have:
\setlength{\arraycolsep}{0.0em}
\begin{eqnarray}
C&\geq &\max_{p^{\ast}\left (u,x\mid s_{1}\right )}\left[ I\left (U;Y,S_{2}\right )-I\left (U;S_{1}\right )\right ]\label{eq.50}\\
&=&\max_{p^{\ast}\left (u\mid x,s_{1}\right )p^{\ast}\left (x\mid s_{1}\right )}\left[ I\left (U;Y,S_{2}\right )-I\left (U;S_{1}\right )\right ]\label{eq.51}\\
&=&\max_{\alpha}\left[ I\left (U;Y,S_{2}\right )-I\left (U;S_{1}\right )\right ]\label{eq.52}
\end{eqnarray}
\setlength{\arraycolsep}{5pt}
where the expression $ I\left (U;Y,S_{2}\right )-I\left (U;S_{1}\right ) $ in (\ref{eq.52}) is calculated for the distributions in $\mathcal{P}_{\rho_{XS_{1}}}^{\ast}$ having properties GC.1-GC.6. Thus, defining $ R\left (\alpha\right )=I\left (U;Y,S_{2}\right )-I\left (U;S_{1}\right )  $, we have:
\begin{equation}
C\geq \max_{\alpha}R\left (\alpha\right )=R\left (\alpha^{\ast}\right ), \label{eq.53}
\end{equation}
therefore $ R\left (\alpha^{\ast}\right )  $ is a lower bound for the channel capacity. To  compute $ R\left (\alpha^{\ast}\right )  $, we write:
\begin{equation}
I\left (U;Y,S_{2}\right )=H\left (U\right )+H\left (Y,S_{2}\right )-H\left (U,Y,S_{2}\right )\label{eq.54}
\end{equation}
and
\begin{equation}
I\left (U;S_{1}\right )=H\left (U\right )+H\left (S_{1}\right )-H\left (U,S_{1}\right ).\label{eq.55}
\end{equation}
For $ H\left (Y,S_{2}\right)$ we have:
\begin{equation}
H\left (Y,S_{2}\right)=\dfrac{1}{2}\log\left (\left (2\pi e\right )^{2}\det\left (cov\left (Y,S_{2}\right )\right )\right )\label{eq.56}
\end{equation}
where 
\begin{equation}
cov\left (Y,S_{2}\right )=\left [e_{ij}\right ]_{2\times 2}\label{eq.57}
\end{equation}
and
\setlength{\arraycolsep}{0.0em}
\begin{equation}
\left. 
\begin{array}{rl}
e_{11}&=P+Q_{1}+Q_{2}+N+2A_{1}+2A_{2}+2B+2L_{0}+2L_{1}+2L_{2},\\
e_{12}&=e_{21}=A_{2}+B+Q_{2}+L_{2}\quad\text{and}  \quad e_{22}=Q_{2}
\end{array}
\right\rbrace \label{eq.58}
\end{equation}
\setlength{\arraycolsep}{5pt}
where the $P$, $Q_{i}$'s, $N$, $A_{i}$'s, $L_{i}$'s and $B$ are defined in previous section. 
Therefore
\begin{equation}
\det\left (cov\left (Y,S_{2}\right )\right )=d_{Q_{1}N}+d_{PN}+d_{PQ_{1}}+2d_{L_{0}L_{1}}-2d_{PL_{1}}-2d_{Q_{1}L_{0}}, \label{eq.60}
\end{equation}
where the terms are defined in (\ref{eq.300}).\\
For $ H\left (U,Y,S_{2}\right ) $ we have:
\begin{equation}
H\left (U,Y,S_{2}\right )=\dfrac{1}{2}\log\left (\left (2\pi e\right )^{3}\det\left (cov\left (U,Y,S_{2}\right )\right )\right )\label{eq.61}
\end{equation}
where
\begin{equation}
cov\left (U,Y,S_{2}\right )=\left [e_{ij}\right ]_{3\times 3}\label{eq.62}
\end{equation}
and
\setlength{\arraycolsep}{0.0em}
\begin{equation}
\left. 
\begin{array}{rl}
e_{11}&=P+\alpha^{2}Q_{1}+2\alpha A_{1},\\
e_{12}&=e_{21}=P+\left (\alpha +1\right )A_{1}+\alpha Q_{1}+\alpha B+\alpha L_{1}+A_{2}+L_{0},\\
e_{13}&=e_{31}=\alpha B+A_{2},\\
e_{22}&=P+Q_{1}+Q_{2}+N+2A_{1}+2A_{2}+2B+2L_{0}+2L_{1}+2L_{2},\\
e_{23}&=e_{32}=A_{2}+B+Q_{2}+L_{2}\quad \text{and}\quad e_{33}=Q_{2}
\end{array}
\right\rbrace \label{eq.63} 
\end{equation}
\setlength{\arraycolsep}{5pt}
after some manipulations we have:
\setlength{\arraycolsep}{0.0em}
\begin{equation}
\det (cov (U,Y,S_{2}))= \left (\alpha -1\right )^{2}d_{N}+\alpha ^{2}d_{P}+2\alpha \left (\alpha -1\right )d_{L_{0}}+2\alpha d_{A_{1}}+2\left (\alpha -1\right )d_{L_{1}}+d_{Q_{1}} \label{eq.68} 
\end{equation}
\setlength{\arraycolsep}{5pt}
For $ H\left (S_{1}\right ) $ and $ H\left (U,S_{1}\right ) $ we have:
\begin{equation}
H\left (S_{1}\right )=\dfrac{1}{2}\log\left (\left (2\pi e\right )Q_{1}\right ).\label{eq.69}
\end{equation}
\begin{equation}
H\left (U,S_{1}\right )=\dfrac{1}{2}\log \left (\left (2\pi e\right )^{2}\det\left (cov\left (U,S_{1}\right )\right )\right )\label{eq.70}
\end{equation}
where 
\begin{equation}
cov\left (U,S_{1}\right )=\begin{bmatrix}
\alpha^{2}Q_{1}+P+2\alpha A_{1}&\alpha Q_{1}+A_{1}\\
\alpha Q_{1}+A_{1}&Q_{1}
\end{bmatrix}\label{eq.71}
\end{equation}
and the determinant of this matrix is:
\begin{equation}
\det\left (cov\left (U,S_{1}\right )\right )=d_{Q_{2}N}.\label{eq.72}
\end{equation}
Substituting (\ref{eq.56}), (\ref{eq.61}), (\ref{eq.69}) and (\ref{eq.70}) in (\ref{eq.54}) and (\ref{eq.55}), we obtain:
\setlength{\arraycolsep}{0.0em}
\begin{eqnarray}
R\left(\alpha\right)=\dfrac{1}{2}\log \left (  \dfrac{d_{Q_{2}N}\bigg[ d_{Q_{1}N}+d_{PN}+d_{PQ_{1}}+2d_{L_{0}L_{1}}-2d_{PL_{1}}-2d_{Q_{1}L_{0}}  \bigg]}{Q_{1}\bigg [   \left (\alpha -1\right )^{2}d_{N}+\alpha ^{2}d_{P}+2\alpha \left (\alpha -1\right )d_{L_{0}}+2\alpha d_{A_{1}}+2\left (\alpha -1\right )d_{L_{1}}+d_{Q_{1}} \bigg ]}       \right ).\label{eq.47}
\end{eqnarray}
\setlength{\arraycolsep}{5pt}
The optimum value of $\alpha$ corresponding to maximum of $R(\alpha)$ is easily obtained as:
\begin{equation}
\alpha^{\ast}=\dfrac{\left (d_{N}+d_{L_{0}}\right )-\left (d_{A_{1}}+d_{L_{1}}\right ) }{d_{N}+d_{P}+2d_{L_{0}}} .\label{eq.48}
\end{equation}
Substituting $\alpha^{\ast}$ from (\ref{eq.48}) into (\ref{eq.47}) and using the equations (\ref{eq.300}), (\ref{eq.34+2})-(\ref{eq.34+5}) and (\ref{eq.43+1}) we finally conclude that $R\left (\alpha^{\ast}\right )$ equals $ R_{G}$ in (\ref{eq.45}). Therefore $ R_{G}$ in (\ref{eq.45}) is a lower bound for the capacity of the channel defined by properties GC.1-GC.4 in \ref{subsec.definition} (details of computations are omitted for the brevity).\qed

\subsection{Appendix B.}
\paragraph*{The proof of Lemma 2} For all distributions $ p\left(y,x,u,s_{1},s_{2}\right)$ in $\mathcal{P}_{\rho_{XS_{1}}}$ defined by properties GC.1-GC.4, we have:
\setlength{\arraycolsep}{0.0em}
\begin{eqnarray}
I\left( U;Y,S_{2}\right)-I\left( U;S_{1}\right)&=&-H\left( U\mid Y,S_{2}\right) + H\left(U\mid S_{1} \right)    \label{eq.74}\\
&\leq & -H\left(U\mid Y,S_{1},S_{2}\right) + H\left(U\mid S_{1}\right )\label{eq.74+1}\\
&=& -H\left(U\mid Y,S_{1},S_{2}\right) + H\left(U\mid S_{1},S_{2}\right )\label{eq.75}\\
&=&I\left (U;Y\mid S_{1},S_{2} \right )\label{eq.76}\\
&\leq &I\left (X;Y\mid S_{1},S_{2}\right )\label{eq.77}
\end{eqnarray}
\setlength{\arraycolsep}{5pt}
where (\ref{eq.74+1}) follows from the fact that conditioning reduces entropy and (\ref{eq.75}) follows from  Markov  chain  $ S_{2}\rightarrow S_{1}\rightarrow UX $  and (\ref{eq.77})  from Markov chain $ U\rightarrow XS_{1}S_{2}\rightarrow Y $ which are satisfied for any distribution in the form of (\ref{eq.2}), including the distributions in the set $\mathcal{P}_{\rho_{XS_{1}}}$. From (\ref{eq.1}) and (\ref{eq.77}) we can write:
\setlength{\arraycolsep}{0.0em}
\begin{eqnarray}
C&=&\max_{p\left ( u,x\mid s_{1}\right ) } \left [ I\left ( U;Y,S_{2}\right ) -I\left ( U;S_{1}\right )\right ]\label{eq.78}\\
&\leq & \max_{p\left ( x\mid s_{1}\right )}\left [I\left ( X;Y\mid S_{1},S_{2}\right ) \right ].\label{eq.79}
\end{eqnarray}
\setlength{\arraycolsep}{5pt}
From (\ref{eq.79}) it is seen that the capacity of the channel cannot be greater than the capacity when both $ S_{1} $ and $ S_{2} $ are available at both the transmitter and the receiver, which is physically predictable. To compute (\ref{eq.79}) we write:
\setlength{\arraycolsep}{0.0em}
\begin{eqnarray}
I\left ( X;Y\mid S_{1},S_{2}\right )&=&H\left (Y\mid S_{1},S_{2}\right )-H\left (Y\mid X,S_{1},S_{2}\right )\label{eq.81}\\
&=&H\left (X+S_{1}+S_{2}+Z\mid S_{1},S_{2}\right )-H\left (X+S_{1}+S_{2}+Z\mid X,S_{1},S_{2}\right )\label{eq.82}\\
&=&H\left (X+Z\mid S_{1},S_{2}\right )-H\left (Z\mid X,S_{1},S_{2}\right )\label{eq.81+1}\\
&=&H\left (X+Z\mid S_{1},S_{2}\right )-H\left (Z\mid S_{1},S_{2}\right ) \label{eq.81+3} \\
&=&H\left (\left (X+Z\right ),S_{1},S_{2}\right )-H\left (S_{1},S_{2},Z\right ), \label{eq.83} 
\end{eqnarray}
\setlength{\arraycolsep}{5pt}
where (\ref{eq.81+3}) follows from the Markov chain $X \rightarrow \left (S_{1},S_{2}\right )\rightarrow Z$. Hence, the maximum value in (\ref{eq.79}) occurs when $ H\left (\left (X+Z\right ),S_{1},S_{2}\right ) $ is maximum. Since $ S_{1} $, $ S_{2} $ and $ Z $ are Gaussian, the maximum in (\ref{eq.79}) is achieved when  $ \left (X,S_{1},S_{2}\right )$ are jointly Gaussian and $ X $ has its maximum power $ P $, in other words, $ I\left (X;Y\mid S_{1},S_{2}\right ) $ must be computed for distribution $ p^{\ast}\left (y,x,s_{1},s_{2}\right ) $ having the properties GC.1-GC.6. Let $ I^{\ast}\left (X;Y\mid S_{1},S_{2}\right ) $ be the maximum value in (\ref{eq.79}). We have:
\begin{equation}
C\leq I^{\ast}\left (X;Y\mid S_{1},S_{2}\right )\label{eq.84}
\end{equation}
To compute $ I^{\ast}\left (X;Y\mid S_{1},S_{2}\right ) $, we first compute  $ H\left (\left (X+Z\right ),S_{1},S_{2}\right ) $ for distribution $ p^{\ast}\left (y,x,s_{1},s_{2}\right ) $ defined by properties GC.1-GC.6:		
\setlength{\arraycolsep}{0.0em}
\begin{equation}
H\left (\left (X+Z\right ),S_{1},S_{2}\right )=\dfrac{1}{2}\log \left (\left (2\pi e\right )^{3}\det \left (cov \left (\left (X+Z\right ),S_{1},S_{2}\right )\right )\right )\label{eq.85}
\end{equation}
\setlength{\arraycolsep}{5pt}
where
\setlength{\arraycolsep}{0.0em}
\begin{eqnarray}
cov\left (\left (X+Z\right ),S_{1},S_{2}\right )&=&E\left\lbrace \begin{bmatrix}
\left (X+Z\right )^{2}\quad &\quad \left (X+Z\right )S_{1}\quad &\quad \left (X+Z\right )S_{2}\\
\left (X+Z\right )S_{1}\quad &\quad S_{1}^{2}\quad &\quad S_{1}S_{2}\\
\left (X+Z\right )S_{2}\quad &\quad S_{1}S_{2}\quad &\quad S_{1}^{2}
\end{bmatrix}
\right\rbrace \\
&=&\begin{bmatrix}
P+N+2L_{0}\quad &\quad A_{1}+L_{1}\quad &\quad A_{2}+L_{2}\\
A_{1}+L_{1}\quad &\quad Q_{1}\quad &\quad B\\
A_{2}+L_{2}\quad &\quad B\quad &\quad Q_{2}
\end{bmatrix}\label{eq.86}
\end{eqnarray}
\setlength{\arraycolsep}{5pt}
and the determinant:
\begin{equation}
\det\left (cov\left (\left (X+Z\right ),S_{1},S_{2}\right )\right )=d_{N}+2d_{L_{0}}+d_{P},\label{eq.87}
\end{equation}
and the other term in (\ref{eq.83}):
\begin{equation}
H\left (S_{1},S_{2},Z\right )= \dfrac{1}{2}\log\left (\left (2\pi e\right )^{3}d_{P}\right )\label{eq.100} 
\end{equation}
where the terms are defined in (\ref{eq.300}).\\
Substituting (\ref{eq.87}) in (\ref{eq.85}), and from (\ref{eq.100}) we have:
\begin{equation}
I^{\ast}\left (X;Y\mid S_{1},S_{2}\right )=\dfrac{1}{2}\log \left (1+ \dfrac{d_{N}+2d_{L_{0}}}{d_{P}}\right ).\label{eq.89}
\end{equation}
Rewriting (\ref{eq.89}) in terms of $\sigma_{X}$, $\sigma_{S_{1}}$, $\sigma_{S_{2}}$, $\sigma_{Z}$, $\rho_{S_{1}Z}$,$\rho_{S_{2}Z}$, and $\rho_{S_{1}S_{2}}$ using (\ref{eq.34+2})-(\ref{eq.34+5}) and (\ref{eq.300}) and taking into account two Makovity results (\ref{eq.43+1}) and (\ref{eq.300x}), we finally conclude that (details of manipulations are omitted for the brevity):
\begin{equation}
I^{\ast}\left (X;Y\mid S_{1},S_{2}\right )=\dfrac{1}{2}\log\left( 1+\dfrac{P}{N} \dfrac{\left (1-\rho_{XS_{1}}^{2}\right ) \left (1-\rho_{S_{1}S_{2}}^{2}\right )}{\quad d_{P}^{\mathcal{N}}} \right ).\label{eq.95}
\end{equation}
Hence, $C$ in (\ref{eq.95+1}) is an upper bound for the capacity of the channel when we have the Markov chain $X \rightarrow \left (S_{1},S_{2}\right )\rightarrow Z$. \qed

\subsection*{Appendix C.}
\paragraph*{Lemma 3}
\quad Two continuous random variables $X$ and $S$ with probability density functions $f_{X}\left (x\right )$ and $f_{S}\left (s\right )$ can be correlated to each other with a specific correlation coefficient $\rho_{XS}$.
\paragraph*{Proof} \quad Suppose $F_{X}(x)$ and $F_{S}(s)$ are the distribution functions of $f_{X}(x)$ and $f_{S}(s)$ respectively. If $X$ and $S$ are jointly distributed with a joint density function $f_{X,S} \left (x,s \right )$ given below, we prove that the correlation coefficient is $\rho_{XS}$:  
\begin{equation}
f_{X,S}(x,s)=f_{X}(x)f_{S}(s)\left[ 1+\rho\left (2F_{X}(x)-1\right )\left (2F_{S}(s)-1\right )\right] \label{eq.200} 
\end{equation}
in which 
\begin{equation}
\rho = \dfrac{\sigma_{X}\sigma_{S}}{a_{X}a_{S}}\rho_{XS}. \label{eq.203} 
\end{equation}
with 
\begin{equation}
a_{X}=\int_{-\infty}^{+\infty}xf_{X}(x)\left (2F_{X}(x)-1\right )dx \label{eq.201} 
\end{equation}
and
\begin{equation}
a_{S}=\int_{-\infty}^{+\infty}sf_{S}(s)\left (2F_{S}(s)-1\right )ds. \label{eq.202} 
\end{equation}
First we note that (\ref{eq.200}) is a joint density function with marginal densities $f_{X}(x)$ and $f_{S}(s)$ \cite[p.176]{papoulis}. Then we need to prove that $E \left\lbrace XS\right\rbrace  =\sigma_{X}\sigma_{S}\rho_{XS} + E\left\lbrace X\right\rbrace E\left\lbrace S\right\rbrace $. From (\ref{eq.200}) we have:
\setlength{\arraycolsep}{0.0em}
\begin{eqnarray}
E\left\lbrace XS\right\rbrace &=& \int_{-\infty}^{+\infty}\int_{-\infty}^{+\infty}xsf_{X,S}(x,s)dxds\\
&=& \int_{-\infty}^{+\infty}\int_{-\infty}^{+\infty}xsf_{X}(x)f_{S}(s)\left[ 1+\rho\left (2F_{X}(x)-1\right )\left (2F_{S}(s)-1\right )\right]dxds\\
&=&E\left\lbrace X\right\rbrace E\left\lbrace S\right\rbrace + \rho a_{X}a_{S}
\end{eqnarray}
\setlength{\arraycolsep}{5pt}
To complete the proof, we need to show that $a_{X}$ and $a_{S}$ in (\ref{eq.201}) and (\ref{eq.202}) exist and have nonzero values. We can show that:
\begin{equation}
\int_{-\infty}^{+\infty}F_{X}(x)\left (1-F_{X}(x)\right )dx= a_{X} + \bigg[ xF_{X}(x)(1-F_{X}(x))\bigg] _{-\infty}^{+\infty}.\label{eq.203+1}   
\end{equation}
The second expression in the right hand side of (\ref{eq.203+1}) is equal to zero because $F_{X}(\pm\infty)\left (1-F_{X}(\pm\infty)\right )$   \textit{is exactly equal} to zero by definition. The integrand at the left hand side of (\ref{eq.203+1}) is a positive and continuous function of $x$ and therefore the integral exists and has nonzero positive value. So $a_{X}$ exists and is nonzero. The same argument is valid for $a_{S}$. \qed  \\

\subsection*{Appendix D.}
\paragraph*{Lemma 4}
\quad Consider three zero mean random variables $ \left(X,S_{1},S_{2}\right)  $ with covariance matrix $ \boldsymbol{K} $ as: 
\begin{eqnarray}
\boldsymbol{K}{}={}E\left\lbrace   \begin{bmatrix}
X^{2} & XS_{1}& XS_{2}\\
XS_{1} & S_{1}^{2} & S_{1}S_{2}\\
XS_{2} &  S_{1}S_{2} & S_{2}^{2}
\end{bmatrix}\right\rbrace  \nonumber \\
{}={}\begin{bmatrix}
\sigma_{X}^{2}&\sigma_{X}\sigma_{S_{1}}\rho_{XS_{1}}&\sigma_{X}\sigma_{S_{2}}\rho_{XS_{2}}\\
\sigma_{X}\sigma_{S_{1}}\rho_{XS_{1}}&\sigma_{S_{1}}^{2}&\sigma_{S_{1}}\sigma_{S_{2}}\rho_{S_{1}S_{2}}\\
\sigma_{X}\sigma_{S_{2}}\rho_{XS_{2}}&\sigma_{S_{1}}\sigma_{S_{2}}\rho_{S_{1}S_{2}}&\sigma_{S_{2}}^{2}
\end{bmatrix}\label{eq.224}
\end{eqnarray}
Suppose $ \left(S_{1},S_{2} \right)  $ are \textit{jointly Gaussian} random variables. Then, if $  \left(X,S_{1},S_{2} \right) $ form Markov chain  $ S_{2}\rightarrow S_{1}\rightarrow X $, \textit{(even if X is not Gaussian)} we have:
\begin{equation}
\rho_{XS_{2}}=\rho_{XS_{1}}\rho_{S_{1}S_{2}}\label{eq.225}
\end{equation}
or equivalently: 
\begin{equation}
E\left\lbrace S_{1}^{2}\right\rbrace E\left\lbrace XS_{2}\right\rbrace =E\left\lbrace XS_{1}\right\rbrace E\left\lbrace S_{1}S_{2}\right\rbrace \label{eq.226} 
\end{equation}

\paragraph*{Proof} we can write:
\setlength{\arraycolsep}{0.0em}
\begin{eqnarray}
\rho_{XS_{2}}&{}={}&\dfrac{E\left\lbrace XS_{2}\right\rbrace }{\sigma_{X}\sigma_{S_{2}}}=
\dfrac{E\left\lbrace E\left\lbrace XS_{2}\mid S_{1}\right\rbrace \right\rbrace }{\sigma_{X}\sigma_{S_{2}}}\label{eq.227}\\
&{}={}&\dfrac{E\left\lbrace E\left\lbrace X\mid S_{1}\right\rbrace E\left\lbrace S_{2} \mid S_{1}\right\rbrace  \right\rbrace} {\sigma_{X}\sigma_{S_{2}}}\label{eq.228}\\
&{}={}&\dfrac{\rho_{S_{1}S_{2}}}{\sigma_{X}\sigma_{S_{1}}}E\left\lbrace S_{1}E\left\lbrace X\mid S_{1}\right\rbrace  \right\rbrace \label{eq.229}\\
&{}={}&\dfrac{\rho_{S_{1}S_{2}}}{\sigma_{X}\sigma_{S_{1}}}E\left\lbrace XS_{1}\right\rbrace \label{eq.230} \\
&{}={}&\rho_{XS_{1}}\rho_{S_{1}S_{2}}\label{eq.231}
\end{eqnarray}
\setlength{\arraycolsep}{5pt}where (\ref{eq.228}) follows from the Markov chain $ S_{2}\rightarrow S_{1}\rightarrow X $ and (\ref{eq.229}) follows from Gaussianness of $ \left( S_{1},S_{2}\right)  $ and the fact that $ E\left\lbrace S_{2}\mid S_{1}\right\rbrace=\frac{\sigma_{S_{2}}\rho_{S_{1}S_{2}}}{\sigma_{S_{1}}}S_{1}  $ and (\ref{eq.230}) follows from the general rule that for random variables $ A $ and $ B $ we have $E\left\lbrace g_{1}\left(A \right)  g_{2}\left(B \right)\right\rbrace =E\left\lbrace g_{1}\left(A \right) E\left\lbrace g_{2}\left(B \right)\mid A\right\rbrace \right\rbrace$ \cite[p.234]{papoulis}. \qed



\end{document}